\newcommand{\be}{\begin{equation}}
\newcommand{\ee}{\end{equation}}
\newcommand{\bea}{\begin{eqnarray}}
\newcommand{\eea}{\end{eqnarray}}
\newcommand{\ba}[1]{\begin{array}{#1}}
\newcommand{\ea}{\end{array}}

\documentclass[twocolumn, prl, amssymb,aps,showpacs,twocolumn,amsmath,showkeys,floatfix]{revtex4-1}
\usepackage{times}
\usepackage{amssymb}
\usepackage{amsmath}
\usepackage{commath}
\usepackage{mathrsfs}
\usepackage{graphicx}
\usepackage{epsfig}
\usepackage{dcolumn}
\usepackage{color}
\usepackage{bm}

\begin{document}

\title{Interfacing superconducting qubits and single optical photons using molecules in waveguides}
\author{Sumanta Das$^{1}$, Vincent E. Elfving$^{1}$, Sanli Faez$^{2}$ and Anders S. S\o rensen$^{1}$}
\affiliation{$^{1}$Niels Bohr Institute, University of Copenhagen, Blegdamsvej 17, 2100 Copenhagen \O, Denmark\\
$^{2}$Debye Institute for Nanomaterials Science and Center for Extreme Matter and Emergent Phenomena, 
Utrecht University, 3584 CC Utrecht, The Netherlands}
\date{\today}

\begin{abstract}
We propose an efficient light-matter interface at optical frequencies between a single photon and a superconducting qubit. The desired interface is based on a hybrid architecture composed of an organic molecule embedded inside an optical waveguide and electrically coupled to a superconducting qubit placed near the outside surface of the waveguide. We show that high fidelity, photon-mediated, entanglement between distant superconducting qubits can be achieved with incident pulses at the single photon level. Such a low light level is highly desirable for achieving a coherent optical interface with superconducting qubit, since it  minimizes decoherence arising from the absorption of light.										
\end{abstract}

\maketitle

Rapid progress in engineering and control of their physical properties, have made superconducting (SC) qubits, one of the most promising candidates for future quantum processors \cite{Scho08, Dev13, Martinis15, Gam15}. If such processors are connected together into a quantum internet \cite{Kim08}, it would allow immense applications ranging from secure communication over long distances \cite{Briegel98,San11, Sre16} to distributed quantum computation \cite{Cirac99, Crep02, Crep06} and advanced protocols for distributed sensing and atomic clocks \cite{Komar13}. Quantum communication over long distances can, however, only be accomplished through optical means making it a necessity to build light-matter interfaces at optical frequencies \cite{Kim08, Gisin}. This has stimulated immense interest in devising ways of efficiently coupling optical photons to SC systems \cite{And04, Rab06, And06, Wall09, Mar10, Taylor11, vitali12, Tian12, clerk12, Gis13, Bri14, Xia14, Pri14, Will14, Duan15, Ham15}. Tremendous success have been achieved in coupling photons to SC qubit at microwave frequencies \cite{Wall04, Gre14}, while in the optical domain, only limited indirect coupling has been achieved using transducers \cite{bochmann13, Andrews14, Bagchi14}. Coherent coupling of quantum fields at optical frequencies to a SC system thus remains an outstanding challenge. A principle obstacle to this is the large mismatch between the energy scales of an optical photon ($\sim 1$ eV) and a SC qubit  ($\sim 100~\mu$eV) \cite{Wall04} making the absorption of even a single optical photon a major disturbance for a SC system. In fact such effects are used in SC detectors for detection of optical photons  \cite{Nat12}. To suppress such disturbances it is therefore highly desirable to keep the number of optical photons to a minimum.

\begin{figure}[!h]
   \begin{center}
   \begin{tabular}{c}
   \includegraphics[height = 6 cm, width = 8.5 cm]{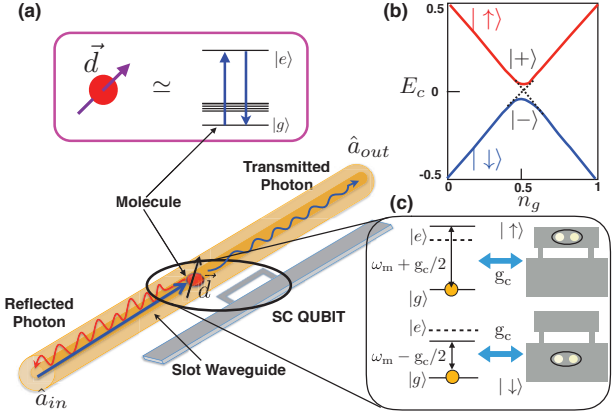}
   \end{tabular}
   \end{center}
   \caption[example] 
   { \label{fig1} 
Schematic of the hybrid molecule-SC system. (a) An organic molecule located inside the optical waveguide is electrically coupled to a SC via a Stark shift. Incident probe photons in the waveguide are elastically scattered by the optical transition of the molecule. Due to the coupling of the SC qubit and the molecule, reflected and transmitted photons are entangled with the internal state of the qubit. (b) The energy levels of a CPB can be represented by two hybridized levels. (c) Oscillation of the Cooper pair between the SC islands leads to shifts of the molecular resonance.}
\end{figure}

In this letter we propose a scheme to {\it interface optical photons with a SC qubit} at light levels involving only a single or a few photons. To achieve this we introduce a hybrid solid-state architecture depicted in Fig. \ref{fig1}(a) comprising, a molecule embedded in an optical waveguide with a SC qubit fabricated near its surface ($\sim 100 - 500$~nm). In comparison to the magnetic coupling considered previously \cite{ Mar10, Bri14, Xia14, Gre14, Pro13, schu10, Tor08, atac09, ver09, zhu11}, a key feature of our scheme is the \textit{electric coupling} between the molecule and SC qubit. The coupling strength can then be orders of magnitude stronger thus allowing for strong coupling in the system. As the SC qubit we consider a Cooper pair box (CPB) where the two quantum states are defined by a single Cooper pair being on each of two superconducting islands. As the Cooper pair oscillates between the islands, it generates a variation in the electric field at the molecule. If the molecule has a large difference in the dipole moment between its ground and excited states, the electric field variation will lead to different Stark shifts of the energy levels (Fig.\ref{fig1} c). This leads to a sizeable shift of the resonance frequency of the molecule, which can be larger than its linewidth, leading to coherent coupling between the molecule and the qubit. Since the molecule is embedded in a waveguide, the shift can lead to measurable effects even for light pulses containing few photons. This is a major advantage over existing hybrid proposals that requires strong optical fields \cite{And04, Dev13, Pri14, Duan15, Gre14, Pro13, schu10, Tor08, atac09, ver09, zhu11,Blum15}, which will lead to decoherence due to quasiparticles created by photon absorption \cite{Cat11}. 

We show how the achieved light-matter interface allows efficient optical readout of a CPB qubit. Furthermore, we present a detailed scheme for photon-mediated entanglement between two distant SC qubits using hybrids with two molecules at each site. The dipole-dipole interaction between these molecules leads to flip-flop processes between them. This induces an oscillating electric field which can drive a resonant transition in the SC qubit. In total this leads to a Raman process, where the emission of an optical Stokes photon is correlated with the excitation of the SC qubit. Combining the output of two such process at a beam splitter and conditioning on a click in a detector allows for long distance entanglement between the SC qubits. This opens the possibility of connecting distant SC quantum computers in a large scale quantum network through teleportation.

The key element in the interface is envisioned to be an organic dye molecule. Such molecules can have large differential Stark shift corresponding to a difference in the static dipole moment of $~1$ Debye between the ground and excited state \cite{Orrit92, Mich99} and can be embedded in optical waveguides \cite{Vahid12, San14, Gaio16, Skoff16}. For organic molecules all the desired properties have thus been demonstrated experimentally, but the molecule could be replaced by any emitter with similar properties. Placing an ideal two level emitter in an optical waveguide, in principle, allows for coupling efficiencies to optical photons of more than $95 \%$ \cite{ Alm04,Qua09}. In practice a coupling efficiency of $10\%$ has been measured \cite{San14}. For the applications proposed in this letter, we show that this is sufficient to achieve operations with few photons per pulse $(\lesssim 1)$.

A CPB resembles a two level system (Fig.\ref{fig1} b) and can be coherently manipulated at temperatures $\leq 100$ mK \cite{Mak01, Dev98, Vion02}. The Hamiltonian of this system can be written as $\mathcal{H}_\text{cp} = -\frac{1}{2}(\chi_1\eta_{\text{z}}+\chi_2\eta_{\text{x}})$, where $\chi_{1,2}$ can be externally controlled while $\eta_{\text{z}},\eta_{\text{x}}$ are the Pauli spin-$1/2$ operators defined in the spin basis $\{|\uparrow\rangle,|\downarrow\rangle\}$ corresponding to distinct charge states \cite{Mak01}. The interaction Hamiltonian $\mathcal{H}_{\text{I}}$ of the molecule-qubit hybrid governing the coherent dynamics can be written in the form \cite{supp}    
\bea
\label{eq:Hint}
&&\mathcal{H}_\text{I}=\frac{\hbar \text{g}_{\text{m}}}{2}\sigma^{+}\hat{a}e^{-i\omega_{\text{p}}\text{t}}+\text{H.c.}+\frac{1}{4}\hbar\text{g}_\text{c}\eta_\text{z}\otimes\left(\sigma^{\text{z}}+\mathbb{I}\right).
\eea 
Here the first term and its Hermitian conjugate correspond to the light-molecule interaction, while the last term is the molecule-CPB interaction. The operators $\sigma^{\text{z}}, \sigma^{\pm}$ are the standard dipole transition operators for a two level system and $\hat{a}$ is the field operator of the incoming photon pulse \cite{book} of central frequency $\omega_{\text{p}}$. The incoming light couples to the molecule with a strength $\text{g}_{\text{m}} = \vec{\wp}_{\text{eg}}\cdot\vec{\mathcal{F}}/\hbar$, where $\vec{\wp}_{\text{eg}}$ is the dipole moment of the optical transition $|e\rangle \leftrightarrow |g\rangle$ in the molecule and $\vec{\mathcal{F}}$ the mode function of the incoming photon in the one-dimensional waveguide. Furthermore, $\text{g}_\text{c} = \Delta\vec{\wp}_{\text{c}}\cdot\Delta\vec{\mathcal{E}}/\hbar$~is the molecule-CPB coupling strength, where $\Delta\vec{\wp}_\text{c}$ is the difference in the static dipole moments between the excited and ground manifold of the molecule, while $\Delta\vec{\mathcal{E}}$ is the electrostatic field variation as seen by the molecule due to the tunnelling of a single Cooper pair. 

We first outline a recipe for detecting the qubit's state by optical photons in a scheme reminiscent of qubit readout in the dispersive regime of circuit QED \cite{Blais04}. We assume that the CPB is operated at a gate voltage away from the charge degeneracy point $\chi_1\gg\chi_2$, i.e. in the linear regime of Fig. \ref{fig1} (b). Working in this regime, the eigenstates of the qubit Hamiltonian $\mathcal{H}_\text{cp}$ are the $\eta^\text{z}$ eigenstates which are (first order) sensitive to the interaction. The CPB-molecule interaction Hamiltonian $(\ref{eq:Hint})$ reveals that, the state of the qubit shifts the excited state of the molecule by $\pm\frac{1}{2}\hbar\text{g}_\text{c}$, compared to the unperturbed resonance at $\omega_\text{m}$ (Fig. \ref{fig1} c) making it sensitive to the qubit state. The molecular resonance line is thus split into two, corresponding to the two qubit states $\{|\uparrow\rangle, |\downarrow\rangle\}$, with the splitting given by the molecule-qubit coupling $\text{g}_\text{c}$. We can therefore determine the qubit state by studying the scattering of an incoming photons and measuring whether they are transmitted or reflected. 
 
Considering a small CPB \cite{Vion02} the coupling can be estimated from the field of a point charge sitting at the edge of a semi-infinite medium. This gives $|\Delta\vec{\mathcal{E}}| \simeq (7 - 19)$ kV/m, at the location of the molecule in a polyethylene waveguide of permitivity $\sim 2.3$ \cite{Wirtz06}, due to presence of a Cooper pair on an island situated about $\sim (500 - 300)$ nm away \cite{Sanprb14}. However in reality, due to the size of the qubit and the composition of the waveguide the electric field is smaller. We find from explicit numerical simulation a value of $|\Delta\vec{\mathcal{E}}| \simeq (4.5 - 16)$ kV/m for an island situated $\sim (500 - 125)$ nm away from the surface of the waveguide \cite{supp}. Hence, for $|\Delta\vec{\wp}_{\text{c}}| = 1$ D \cite{Orrit92, Mich99}, we can obtain a coupling strength of $\text{g}_\text{c} \sim (2\pi)\times (25 - 80)$ MHz. As organic molecules typically have optical transitions with narrow linewidths $\gamma\sim (2\pi)\times20$ MHz \cite{Orrit92, Vahid12}, we can achieve a strong coupling regime ($\text{g}_\text{c}> \gamma$) where the molecular line splitting exceeds the linewidth. Thus the two internal states of the qubit can be distinguished by sending in a pulse resonant with one of the resonance peaks and measuring whether photons are reflected. At resonance we evaluate the reflection probability \cite{supp} to be $(\gamma_{1D}/\gamma)^2$, where $\gamma_{1D}$ is the decay rate into the $1$D waveguide. Hence we can distinguish the two states by sending in $(\gamma/\gamma_{1D})^2\sim 100$ photons for the experimentally observed efficiency of $\gamma_{1D}/\gamma= 0.1$ \cite{San14}. 

To achieve a coherent interface we chose the qubit to be at the so called sweet spot ($\chi_{1} = 0$) where the energy of the eigenstates $|\pm\rangle = (|\uparrow\rangle\pm|\downarrow\rangle)/\sqrt{2}$, of the CPB Hamiltonian are first order insensitive to charge noise (Fig. 1 b) and thus can have long coherence time \cite{Ith05,Hou09}. To be able to work with a few photons we consider, the {\it Raman scattering scheme} shown in Fig. \ref{fig2} (a). This is realized, by having two organic molecules with properties as above and with optical transitions of nearly the same frequencies, e.g., by tuning them into resonance using an external field. The molecules are assumed to have a separation less than the optical wavelength, and thus couple to each other via near field optical dipolar interaction \cite{Hett02}. Furthermore, we assume a reflector at one end of the waveguide such that the waveguide is single sided to maximize the collection of Raman photons.

The interaction Hamiltonian $\mathcal{H}_{I}$ comprise the dipolar coupling Hamiltonian $\mathcal{H}_{\text{dd}} = \hbar\text{V}(\sigma^{+}_{1}\sigma^{-}_{2}+ \sigma^{+}_{2}\sigma^{-}_{1})$ of strength $V$, and the Hamiltonian
\bea
\label{eq:HRint}
\mathcal{H}_{\text{c}} = \sum_{j}\left[\frac{\hbar\text{g}_{\text{m}_j}}{2}\sigma^{+}_{j}\hat{a}e^{-i\omega_{\text{p}}t}+\text{H.c.}+\frac{\hbar\text{g}_{\text{c}_j}}{4}\eta_{\text{z}}\otimes\left(\sigma^{\text{z}}_j+\mathbb{I}\right)\right],\nonumber\\
\eea
where $\text{g}_{\text{m}_j}$ and $\text{g}_{\text{c}_j}$ for $(j = 1, 2)$ corresponds to the coupling strength of the incoming light and CPB to the molecules respectively. The strong dipole-dipole interaction $\mathcal{H}_\text{dd}$ can be diagonalized to form two dressed state $|S\rangle$ and $|A\rangle$ which are split by $2\mathcal{V} = \sqrt{4\text{V}^{2}+\delta^{2}_{0}}$ and have an electrical dipole transition between them. Using an external field to vary the difference in the molecular energies $\delta_{0} = (\omega_{m_1}-\omega_{m_2})$, the transition between the dressed states can be brought into resonance with the qubit transition, $2\mathcal{V} = \omega_\text{q}$. This resonance condition allows the exchange of energy between the qubit and the excited manifold of the molecules through the interaction of the dipoles with the charge fluctuations in the CPB. This enables the Raman transition $|g,-\rangle\rightarrow|S,-\rangle\rightarrow|A,+\rangle\rightarrow|g,+\rangle$ (Fig. \ref{fig2} a). Here the molecular system starts and ends in the joint ground state $|g\rangle = |g_1,g_2\rangle$ while the qubit is flipped from state $|-\rangle$ to $|+\rangle$ by the emission of a Stokes photon of frequency $\omega_\text{s} = (\omega_\text{p}-\omega_\text{q})$. The effective coupling constant between the states $|S,-\rangle$ and $|A,+\rangle$ which enables this transition is given by $\mathcal{G} = (\text{g}_{\text{c}_1}-\text{g}_{\text{c}_2})\text{V}/\sqrt{4\text{V}^{2}+\delta^{2}_{0}}$. Using the effective operator formalism \cite{flo12}, we find that at resonance the probability for an incident single photon to induce a Raman scattering into the waveguide, for moderate coupling $\text{g}^{2}_{\text{c}_{1,2}}/\gamma\omega_\text{q} < 1$ is $\mathcal{P}_{R} = \left(\gamma_{1D}/\gamma\right)^{2}\wp_{R}$, where \cite{supp},
\bea
\label{eq:pr}
\wp_{R} =\left(\frac{\delta_{0}}{\omega_\text{q}}\right)^{2}\left(\frac{4\mathcal{G}^{2}}{\Gamma^{2}_s\Gamma^{2}_a/4\gamma^{2}+4\mathcal{G}^{2}}\right).
\eea 
Here $\gamma = \gamma_{1D}+\gamma_{c}+\gamma_{i}$ is the total decay rate of each molecule (assumed identical for the two molecules), $\Gamma_{s} = \gamma + 2\gamma_{c}\text{V}/\omega_\text{q}, \Gamma_{a} = \gamma - 2\gamma_{c}\text{V}/\omega_\text{q}$ are the decay rates of $|S\rangle$ and $|A\rangle$ respectively, $\gamma_{i}$ is the intrinsic decay rate of each molecule while $\gamma_c$ is the collective decay rate of the molecules. In deriving the Raman scattering probability we assumed that the molecules have the same $\gamma_{1D}$ and that they are close enough, that we can ignore phases in the collective decay \cite{Das08} arising from the spatial positions of the molecules. The probability of Raman scattering is much larger than the reverse process $|A,+\rangle \rightarrow |S,-\rangle$, which is suppressed by a factor $(\mathcal{G}\gamma)^2/\omega^{4}_\text{q}$ since it is off resonant \cite{supp}. 
\begin{figure}
\begin{center}
\includegraphics[height=4.4 cm]{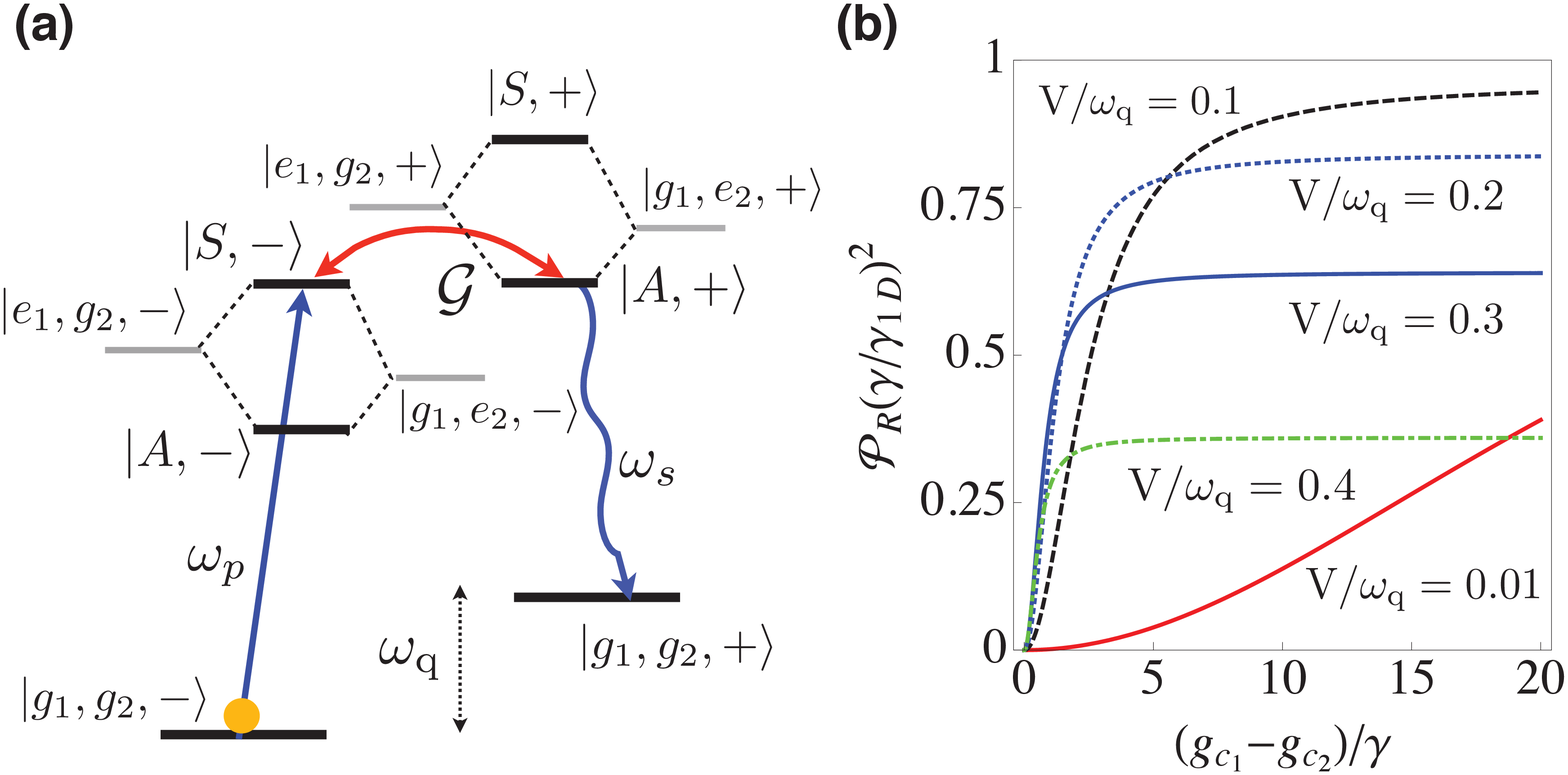}
\caption[example] 
{ \label{fig2} 
(a) Schematics of Raman configuration. The molecular levels $|e_1g_2\rangle$ and $|g_1 e_2\rangle$ are hybridized by the dipole-dipole interaction to form the dressed states $|A\rangle$ and $|S\rangle$, the separation of which is tuned into resonance with the qubit frequency $\omega_\text{q}$. A photon scattered along the transition $|g,-\rangle\rightarrow|S,-\rangle$ is emitted as a Stokes photon along the transition $|A,+\rangle \rightarrow |g,+\rangle$ due to resonant coupling among the states $|S,-\rangle \leftrightarrow |A,+\rangle$. (b) Normalised probability of Raman scattering derived in Eq. (\ref{eq:pr}) for a single incident photon as a function of coupling $(\text{g}_{\text{c}_1}-\text{g}_{\text{c}_2})/\gamma$ for different values of the dipolar coupling strength. Here we have assumed $\gamma_{c}/\gamma_{i} = 1$}
\end{center} 
\end{figure} 

In Fig. \ref{fig2} (b) we plot the Raman probability as function of $\left(\text{g}_{\text{c}_1}-\text{g}_{\text{c}_{2}}\right)/\gamma$ for different ratios of $\text{V}/\omega_{q}$. The results can be understood from the need to have both good hybridization and coupling to the waveguide. For low $\text{V}/\omega_q $ the hybridization of $|e_1g_2\rangle $ and  $|g_2e_1\rangle $ to $|A\rangle$ and $|S\rangle$ is small which limits the coupling, whereas for $\text{V}/\omega_q\rightarrow 1/2$, $\delta_0\rightarrow 0$, $|A\rangle $ becomes a dark state of the coupling to the waveguide so that $\mathcal{P}_R\rightarrow 0$. For $\text{V}/\omega_\text{q} \gtrsim 0.1$, the probability quickly reaches its maximum value even for limited coupling strengths $(\text{g}_{\text{c}_{1}}-\text{g}_{\text{c}_{2}})/\gamma \gtrsim 1$, whereas saturation is slower for weaker dipole coupling due to the lack of hybridization. We find from Fig. \ref{fig2} (b), that for a feasible $\text{V}/\omega_{\text{q}} = 0.2$ and a moderate coupling strength $(\text{g}_{\text{c}_{1}}-\text{g}_{\text{c}_{2}})/\gamma = 4$, the Raman scattering probability is $\mathcal{P}_{R} \simeq 0.77 \times (\gamma_{1D}/\gamma)^{2}$ (note that since the $g_{c_1}$ and $g_{c_2}$ can have opposite signs, their difference can exceed their individual values). These parameters will be used for all numerical examples below. The value of $\mathcal{P}_{R}$, is close to its upper limit of $(\gamma_{1D}/\gamma)^2$ set by the necessity of having both waveguide absorption and emission by the molecule. Furthermore, the Raman probability is not very sensitive to the precise value of the dipole coupling making it attractive even for randomly placed molecules.  

\begin{figure}[!h]
\begin{center}
\includegraphics[width = 8 cm, height= 7 cm]{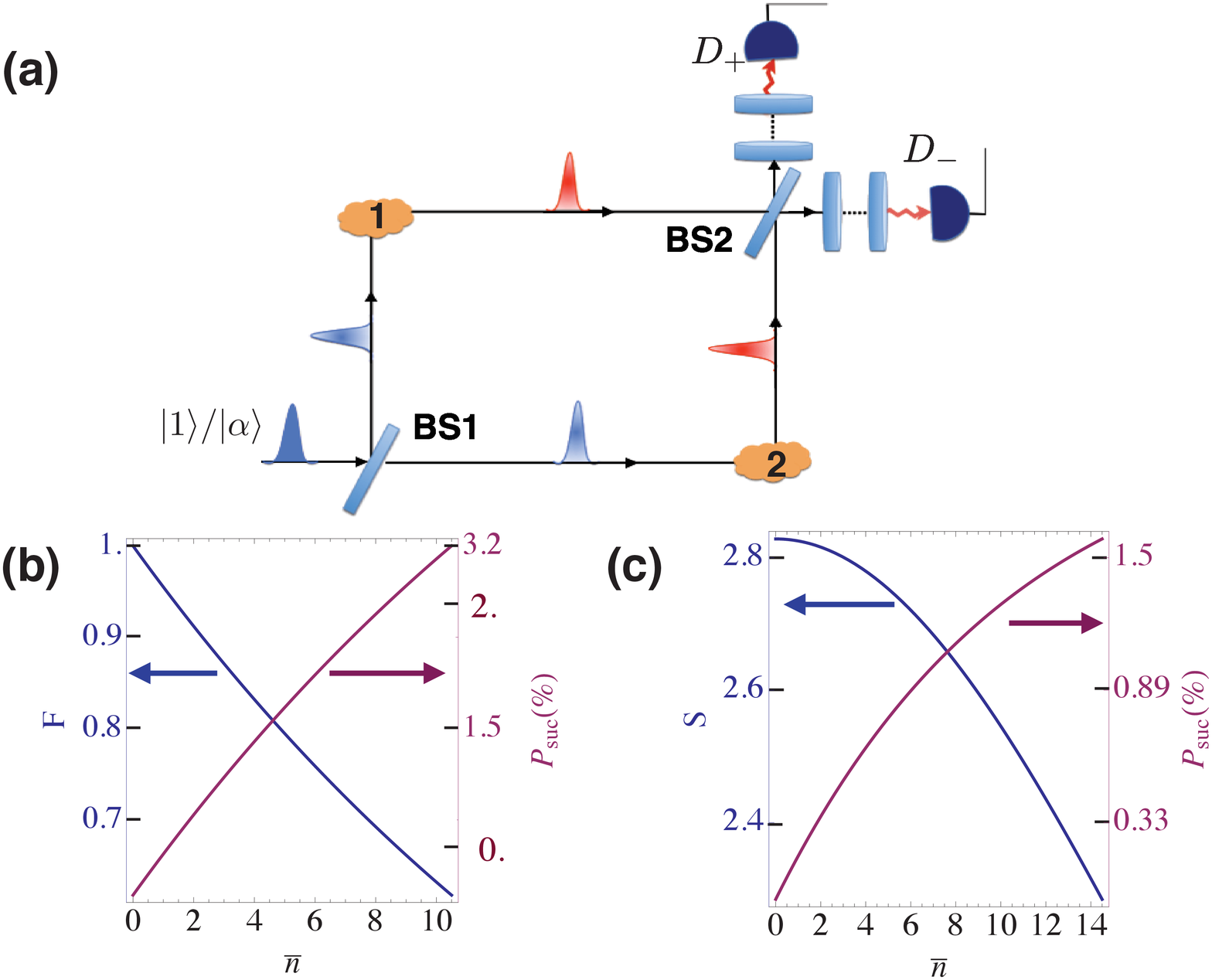}
\caption{\label{fig22} (a) Interferometric scheme to generate maximally entangled Bell state $|\Psi_{+}\rangle$ between two SC qubits using a single photon $|1\rangle$ or coherent state $|\alpha\rangle$ as input. Generation of entanglement is conditioned on a click of detector $D_{\pm}$. (b) Fidelity $F$ and success probability $P^{(c)}_\text{suc}$ for Bell state generation with a pulse with $\bar{n}$ photons on average. (c) Bell parameter $S$ and success probability $P^{(c)}_\text{suc}$ for an entangled state between a single qubit and a photon, for an incident pulse of $\bar{n}$ photons. For all the plots we have assumed $\gamma_{1D}/\gamma = 0.1$, $\eta = 50\%$, $(\text{g}_{\text{c}_{1}}-\text{g}_{\text{c}_{2}})/\gamma = 4$, $\mathcal{P}_{R} = 0.77\times(\gamma_{1D}/\gamma)^{2}$ and $\gamma_{c}/\gamma = \gamma_{i}/\gamma = 0.45$}
\end{center}
\end{figure}

The effective Raman scheme, allows using the interferometric framework \cite{zoll99,Lillian05}, shown schematically in Fig. \ref{fig22} (a) to generate entanglement between distant (e.g. $10$'s kms) SC qubits via the detection of a photon. For this purpose we assume that both hybrids are initialized in state $|g,-\rangle_{1(2)}$. An incident photon pulse (blue in \ref{fig22}(a)) is split by the beam splitter and sent towards the two interfaces, where it can induce Raman transitions.  An outgoing photon (red) is correlated with a transition to $|+\rangle$ in the corresponding qubit. Interfering the outputs on a beam splitter erases the "which way" information about which interface emitted the photon. Hence by conditioning on clicks in detectors $D_\pm$ after frequency filtering out photons, which have not undergone Raman scattering, the qubits are projected into one of the maximally entangled Bell states $|\Psi_{\pm}\rangle = \frac{1}{\sqrt{2}}|g\rangle\left(|-\rangle_{1}|+\rangle_{2}\pm|+\rangle_{1}|-\rangle_{2}\right)$ depending on which detector clicks. With a single incident photons this process creates an ideal Bell states, provided that no other sources of noise are present. 

For simplicity we consider the two hybrids to have equivalent physical properties and work in the limit of moderate coupling $\text{g}^{2}_{\text{c}_{1,2}}/\gamma\omega_{\text{q}} < 1$. Assuming the input pulse to be a single photon, we find that within the present model the process has a fidelity $F = 1$, and a success probability of $P^{(1)}_{\text{suc}} = \eta\mathcal{P}_{R}$, where $\eta$ is the photodetection efficiency of the single photon detectors \cite{supp} [in reality $F < 1$, due to e.g. dephasing of the qubits during the time needed for scattering of photons (see below)]. For the set of parameters used above along with $\gamma_{1D}/\gamma=0.1$, we get $P^{(1)}_{\text{suc}} \simeq 3.8\times10^{-3}$, for $\eta = 50\%$. 

A simpler experimental method is to use a weak coherent state as input. Assuming identical hybrids and an intensity below saturation, we find the conditional fidelity for an incident photon pulse with $\bar{n}$ photons in the lowest order to be $F = 1-\frac{\bar{n}}{2}\left(\mathcal{P}_{R}+\mathcal{P}_{RO}+\mathcal{P}_{D}/4\right)$, with the corresponding success probability \cite{supp}
\bea
\label{eq4}
P^{(c)}_{\text{suc}} = 2P^{(1)}_{\text{suc}}\left\{\frac{1-e^{-\frac{\bar{n}}{2}\left(\mathcal{P}_{R}+\mathcal{P}_{RO}\right)}}{\mathcal{P}_{R}+\mathcal{P}_{RO}}\right\},
\eea
where the probability of Raman scattering to the outside (not into the waveguide) is $\mathcal{P}_{RO} = \left(\frac{\gamma_{1D}}{\gamma}\right)\left\{\left(\frac{\gamma_{c}}{\gamma}\right)\left(\frac{\delta_{0}}{\omega_\text{q}}\right)^{2}+\left(\frac{\gamma_{i}}{\gamma}\right)\left(1+\frac{2V}{\omega_\text{q}}\right)\right\}\left(\frac{2\mathcal{G}^{2}}{\Gamma^{2}_s\Gamma^{2}_a/4\gamma^{2}+4\mathcal{G}^{2}}\right)$ while $\mathcal{P}_{D}$ is the probability of dephasing induced by 
elastic scattering \cite{supp}. As an example, for $\bar{n} = 1.5$ we get a success probability $P^{(c)}_{\text{suc}} \simeq 5.6\times10^{-3}$ for creating a Bell state with fidelity $F \sim 90\%$. Thus for an input coherent state, we gain substantially in experimental simplicity with limited reduction in fidelity. We show in Fig. \ref{fig2} (b), the behavior of $F$ and $P^{(c)}_\text{suc}$ as a function of the mean photon numbers.   

 A first step towards the above entanglement generation scheme can be achieved in a simpler setup involving only a single SC qubit, by replacing the other hybrid in the interferometer in the Fig. \ref{fig22}(a) with a frequency shifter. In this case one can obtain a violation of the CHSH inequality $S \leq 2$ \cite{CHSH1,CHSH2}, where $S$ is the violation parameter, between the qubit and a single photon detected at $D_\pm$  \cite{supp}. In Fig. \ref{fig22}(c) we show the behavior of $S$ and $P^{(c)}_\text{suc}$ as a function of the mean photon number. Using the above parameters we find $S \geq 2.3$ for $\bar{n} = 2$, with a corresponding success probability $P^{(c)}_\text{suc} \simeq 1.5\%$ for $\eta = 50\%$.

So far we have ignored the time $T$ it takes to perform the light scattering. The scattering needs to be completed within the coherence time of the qubit. Since we condition on photon detection, the interaction strength only enters through the success  probability as contained in Eq. (\ref{eq:pr}). The time is therefore only limited by the requirement to be resonant with states of width $\gamma$, implying $\gamma T \gg 1$. With $\gamma= (2\pi) 20$ MHz we can chose a pulse duration $T = 50$ ns. Since CPBs of the type considered here have shown coherence times of $T_{2} = 500$ ns \cite{Wall05} with a Gaussian decay, we estimate a further reduction of the fidelity by $ (T/T_{2})^{2} < 1\%$ due to qubit decoherence \cite{supp}.  

An attractive feature of the Raman scheme is that it relies on an electric coupling at the qubit resonance frequency. This scheme can thus be extended to more advanced qubit designs, such as transmons which are insensitive to low frequency electric noise and thus have very long coherence times \cite{Hou09}. The larger size of these qubits, however, diminish the coupling making it infeasible for the molecular parameters considered here. This could possibly be overcome with qubit designs optimized for this purpose or by using other emitters with larger dipole moments. Alternatively one can envision dedicated CPB qubits acting as communication interfaces for quantum computers based on transmons.

In conclusion we have proposed a novel hybrid system formed by an organic molecule embedded in an optical waveguide and electrically coupled to a SC qubit, that provide a light-matter interface for quantum information transfer over long distances. This could open new directions in quantum communication using SC quantum processors in a network.

\begin{acknowledgments}
S. D., V. E., and A. S. gratefully acknowledge financial support from the European
Union Seventh Framework Programme ERC Grant QIOS (Grant No. 306576) and the
Danish Council for Independent Research (Natural Sciences). S. F. acknowledges
support by the European Research Council, Project No. 279248
\end{acknowledgments}


\renewcommand{\theequation}{S\arabic{equation}}
\renewcommand{\thesection}{S\arabic{section}}
\renewcommand{\thefigure}{S\arabic{figure}}

\newpage
\onecolumngrid

\begin{center}
\Large{Supplementary Material}
\end{center}
\section{System Hamiltonian and State Detection}
Here we give the full system Hamiltonian and describe the details of the qubit detection scheme. The Hamiltonian of the combined molecule-CPB qubit hybrid system can be written as $ \mathcal{H} = \mathcal{H}_{0}+\mathcal{H}_{\text{I}}$, where $\mathcal{H}_{0} = \mathcal{H}_{\text{cp}}+\mathcal{H}_{\text{m}}+\mathcal{H}_{\text{f}}$. Here $\mathcal{H}_{\text{cp}} = \frac{1}{2}\left(\chi_{1}\eta_{\text{z}}+\chi_{2}\eta_{\text{x}}\right)$, is the free energy Hamiltonian of the qubit with $\chi_{1} \propto E_{\text{c}}(1-2n_\text{g})$ and $\chi_{2}\propto E_{\text{J}}$, where $E_{\text{c}}$ and $E_{\text{J}}$ are the Coulomb energy of an extra pair of charge on the island and the Josephson energy respectively. The gate charge $n_{g}$ is defined by $n_{g} = C_{g}V_{g}/2e$, where, $C_{g}$ and $V_{g}$ are the gate capacitance and voltage respectively while $e$ is the charge of an electron. In the charge regime (Josephson junction energy $E_\text{J} \ll E_\text{c}, n_g \neq 1/2$), the qubit states are eigenstate of the $\eta_{\text{z}}$ Hamiltonian and hence the free energy Hamiltonian for the qubit becomes $\mathcal{H}_{\text{q}} = \frac{1}{2}\hbar\omega_{\text{q}}\eta_{\text{z}}$ where $\omega_{\text{q}}$ is the qubit transition frequency $\propto E_{\text{c}}(1-2n_\text{g})/\hbar$. The Hamiltonian $\mathcal{H}_{\text{m}} = \frac{1}{2}\hbar\omega_{\text{m}}\sigma^{z}$ is the free energy Hamiltonian of the molecule with $\omega_{\text{m}}$ being the transition frequency of the optical dipole in the molecule and $\mathcal{H}_{\text{f}} = \sum_{\text{k}}\hbar\omega_{\text{k}}(\hat{a}^{\dagger}_{\text{k}}\hat{a}_{\text{k}}+1/2)$, is the free field Hamiltonian with $\hat{a}_{\text{k}}$ being the field operator of mode $\text{k}$ and frequency $\omega_{\text{k}}$. The interaction Hamiltonian $\mathcal{H}_{\text{I}}$ can be divided into two parts $\mathcal{H}^{\text{I}}_{\text{m-L}}$ and $\mathcal{H}^{\text{I}}_{\text{m-q}}$. The Hamiltonian $\mathcal{H}^{\text{I}}_{\text{m-L}}$, describe the interaction between the incoming photon and the molecule and can be written 
\bea
\label{eq1}
\mathcal{H}^{\text{I}}_{\text{m-L}} & = &\frac{\hbar\text{g}_{\text{m}}}{2}\sigma^{+}\hat{a}e^{i[kr-\omega_{\text{p}}t]}+\frac{\hbar\text{g}_{\text{m}}}{2}\hat{a}^{\dagger}\sigma^{-}e^{-[ikr-\omega_{\text{p}}t]}.
\eea
$\mathcal{H}^{I}_{\text{m-q}}$ is the interaction between the molecule and the SC qubit and has the structure
\bea
\label{eq2}
\mathcal{H}^{\text{I}}_{\text{m-q}} = \frac{\hbar\text{g}_{\text{c}}}{4}\eta_{\text{z}}\otimes\left(\sigma^{\text{z}}+\mathbb{I}\right),
\eea
where $\eta_{\text{z}} = (|\downarrow\rangle\langle\downarrow|-|\uparrow\rangle\langle\uparrow|)$, $\sigma^{\text{z}} = \left(|e\rangle\langle e|-|g\rangle\langle g|\right)$, $\sigma^{+} = |e\rangle\langle g| (\sigma^{-} = [\sigma^{+}]^{\dagger})$ and we have renormalized the couplings as $\text{g}_{\text{m}} = \tilde{\text{g}}_{\text{m}}/\sqrt{v_g}$ and $\text{g}_{\text{c}}=\tilde{\text{g}}_{\text{c}}/\sqrt{v_g}$ such that, the field mode operators becomes $\hat{a} \rightarrow \sqrt{v_g}~\hat{a}$. The bare light-molecule and molecule-qubit couplings $\tilde{\text{g}}_{\text{m,c}}$ are defined explicitly in the main text. 

To investigate the effect of coupling between the molecule and the CPB, and to develop a scheme for the detection of the qubit state, we now study coherent scattering of optical photons from the hybrid system. Due to the light-matter interaction, the scattering maps the qubit state onto the scattered optical photons, and the detection of these then provide information about the qubit state. We here assume the CPB to be operated at a gate voltage away from the sweet spot $(n_g \neq 1/2)$. The qubit levels are then given by the eigenstates $\{|\downarrow\rangle, |\uparrow\rangle\}$ of the operator $\eta^{\text{z}}$. Furthermore, to study the dynamics of the hybrid, we choose a combined molecule-CPB qubit basis $\{|e,\downarrow\rangle, |e,\uparrow\rangle,|g,\downarrow\rangle,|g,\uparrow\rangle \}$ and use an effective operator formalism, where, one eliminates the excited state manifold such that the dynamics involves only the lower states with effective decay rates, detuning and couplings as prescribed in Ref.\cite{flo12}. To study the scattering of photons inside the waveguide, we adopt an input-output formalism in the Heisenberg picture, for the field mode operators 
\bea
\label{eq3}
\hat{a}^{f}_{o}(\text{z},t) & = &\hat{a}^{f}_{\text{in}}(\text{z}-v_{g}t) +i\sum_{\text{mm}'}e^{-i\omega_{\text{mm}'}(\text{z}'-\text{z})/v_{g}}\rho_{\text{mm}'}(t)\left[\zeta^{ff}_{\text{m}'\text{m}}\hat{a}^{f}_{\text{in}}(\text{z}-v_{g}t)+e^{-2ik_{0}\text{z}'}\zeta^{fb}_{\text{m}'\text{m}}\hat{a}^{b}_{\text{in}}(\tilde{\text{z}}+v_{g}t)\right],\\
\label{eq3a}
\hat{a}^{b}_{o}(\text{z},t) & = &\hat{a}^{b}_{\text{in}}(\text{z}+v_{g}t) +i\sum_{\text{mm}'}e^{i\omega_{\text{mm}'}(\text{z}'-\text{z})/v_{g}}\rho_{\text{mm}'}(t)\left[\zeta^{bb}_{\text{m}'\text{m}}\hat{a}^{b}_{\text{in}}(\text{z}+v_{g}t)+e^{2ik_{0}\text{z}'}\zeta^{bf}_{\text{m}'\text{m}}\hat{a}^{f}_{\text{in}}(\tilde{\text{z}}-v_{g}t)\right],
\eea
where $\rho_{\text{mm}'}$ is the density operator involving the ground states of the emitter while the superscripts $f (b)$ stands for the forward (backward) travelling wave, $\hat{a}_{o}$ gives the outgoing photon, $\hat{a}_{\text{in}}$ is the incoming photon annihilation operator, $\tilde{\text{z}} = (2\text{z}'-\text{z})$ while $\text{z}'$ and $\text{z}$ are the position of the scatterer and observation respectively.  For the group velocity of the photon wave-packet inside the waveguide $v_{g}$, similar dispersion in the forward and backward directions is assumed, and $\text{m, m}'$ are the indices corresponding to all possible initial and final states (attained after the scattering) of the scatterer. The scattering co-efficient, $\zeta$ is evaluated to be      
\bea
\label{eq4}
\zeta^{ij}_{\text{m}'\text{m}} = \sum_{\text{ee}'}\left(\sqrt{\frac{\Gamma^{i}_{\text{m}'\text{e}}}{2}}(\mathcal{H}_{\text{nh}})^{-1}_{\text{ee}'}\sqrt{\frac{\Gamma^{j}_{\text{e}'\text{m}}}{2}}~\right),
\eea 
where, $\mathcal{H}_{\text{nh}}$ is a non-Hermitian Hamiltonian defined as $\mathcal{H}_{ee'}-\frac{i}{2}\sum_{k}\mathcal{L}^{\dagger}_{k}\mathcal{L}_{k}$, where $\mathcal{H}_{ee'}$ is part of the Hamiltonian $\mathcal{H}_{0}$ in the excited state manifold and $k$ stands for the different possible decay paths from the excited state manifold. Here $\text{e,e'}$ are indices corresponding to the excited states of the scatterer, while the rate of scattering into the one-dimensional mode of the optical slot waveguide from the transition $|e\rangle \leftrightarrow |m\rangle$ of the scatterer is given by $\Gamma^{i}_{\text{e}'\text{m}} \propto (\text{g}^{i}_{\text{m}}[e'\rightarrow m])^{2}$. Note that the above input-output relation derived by generalization of \cite{flo12} is independent of the kind of scatterers and applies to a multitude of problems involving photon scattering in waveguides \cite{Das16}.

In our coherent scattering scheme, $\text{m} = \text{m}'$ and $\text{e} = \text{e}'$ with $\text{m}\equiv \{|g, \downarrow\rangle,|g, \uparrow\rangle\}$ while $\text{e}\equiv \{|e,\downarrow\rangle,|e,\uparrow\rangle\}$. The reflected photon $a^{r}$ according to the above input-output relation is then 
\bea
\label{eq4a}
\hat{a}_{r}(\text{z},t) =\sum_{m}e^{2ik_{0}\text{z}'}\rho_{mm}(t)\zeta_{\text{m}\text{m}}\hat{a}^{f}_{\text{in}}(\tilde{\text{z}}-v_{g}t),
\eea
where for simplicity we ignore the vacuum noise contribution from modes initially not containing any photons. This is justified since we will consider photo detection, where the vacuum modes never results in clicks in detectors. Mathematically this is accounted for by the normal ordering appearing in the description of the measurement protocols below. In such normally ordered products the contribution from vacuum noise vanishes.
For scattering with a single photon pulse of carrier frequency $\omega_{\text{p}}$, we find that the scattering co-efficient for the transition pathways $|e,\downarrow\rangle \rightarrow |g, \downarrow\rangle \rightarrow |e,\downarrow\rangle$ and $|e,\uparrow\rangle \rightarrow |g, \uparrow\rangle \rightarrow |e,\uparrow\rangle$ to be respectively
\bea
\label{eq5}
\zeta_{g,\downarrow} =\left(\frac{\gamma_{1D}}{2}\right)(\Delta - i\gamma/2-\text{g}_{\text{c}}/2)^{-1},\qquad \mathcal{\zeta}_{g,\uparrow} = \left(\frac{\gamma_{1D}}{2}\right)(\Delta - i\gamma/2+\text{g}_{\text{c}}/2)^{-1}, 
\eea
where the detuning is $\Delta =(\omega_{m}-\omega_{\text{p}})$, $\gamma = \gamma'+\gamma_{1D}$ is total radiative decay rate of the molecular transition with $\gamma'$ being the decay to the surrounding and $\gamma_{1D} = \Gamma_{\text{e}\uparrow\text{m}\uparrow} = \Gamma_{\text{e}\downarrow\text{m}\downarrow}$, that into the one-dimensional waveguide. Furthermore we find $\rho_{mm}(t) = \rho_{mm}(0) = 1$ from the master equation for the density matrix when the hybrid is initially prepared in the state $|g, \downarrow\rangle$ or $|g, \uparrow\rangle$. The input-output relation then gives the scattered photon depending on the initial state of the qubit $|\downarrow\rangle$ or $|\uparrow\rangle$. From Eq. (\ref{eq5}) we find that resonant scattering occurs by satisfying the resonance conditions $\Delta = \mp \text{g}_\text{c}/2$.  

\section{Estimation of the qubit-molecule coupling strength ($\text{g}_{\text{c}})$}
\begin{figure}[!h]
\begin{tabular}{ccc}
		\includegraphics{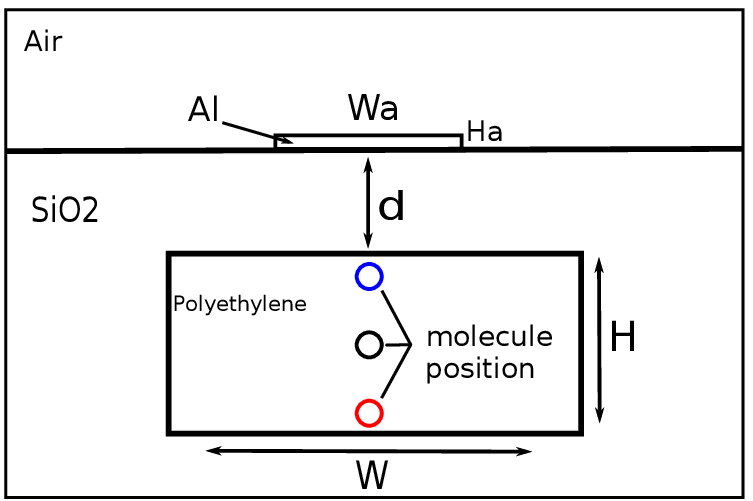} & & \includegraphics[height = 5.3 cm, width= 8cm]{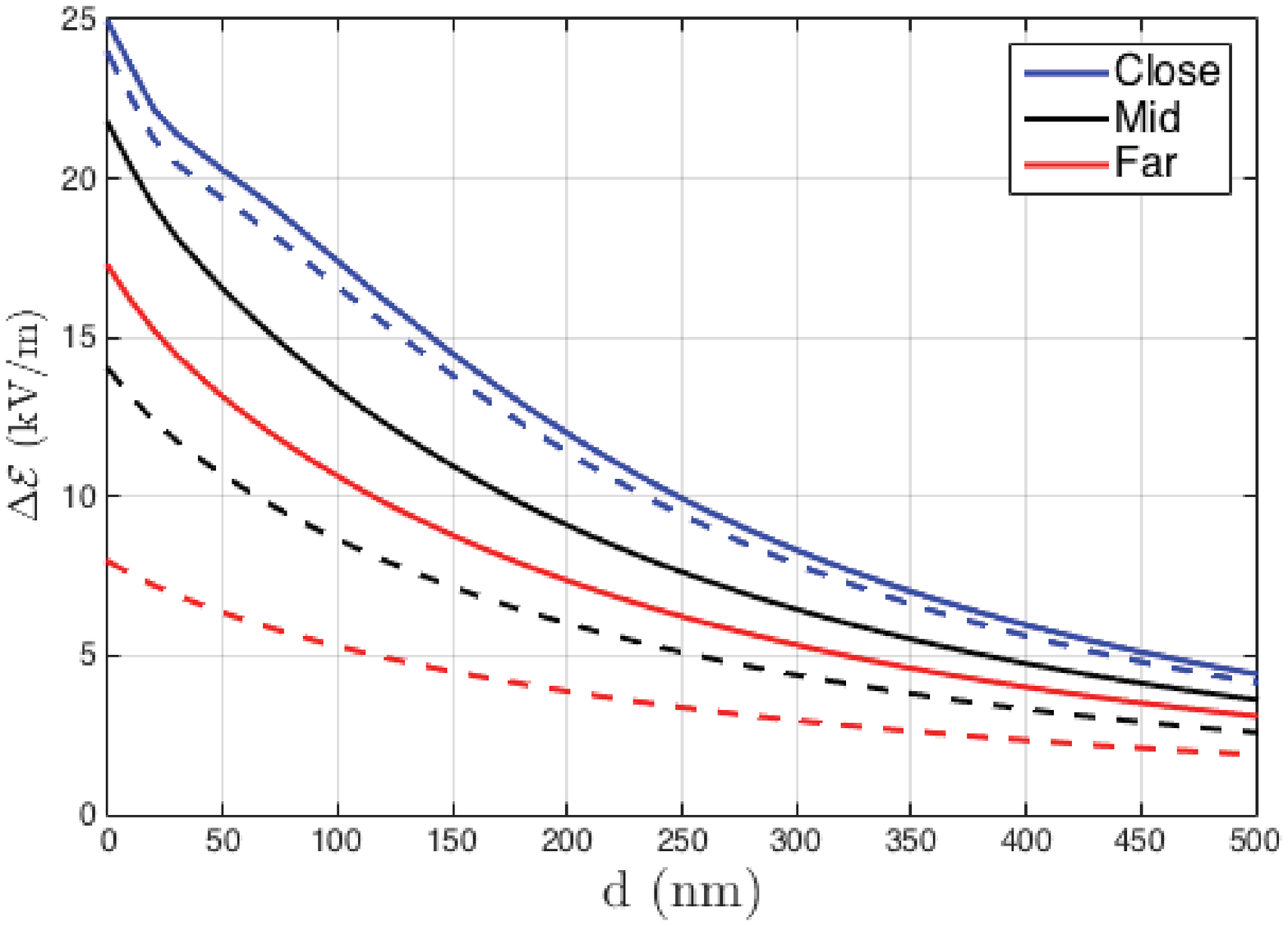}\\
		(a) & & (b)
\end{tabular}
\caption{(a) Schematic diagram showing a polyethylene waveguide of height H and width W in a $\mathrm{SiO_2}$ substrate with an aluminum (Al ) superconducting charge qubit island defined on top. A single Cooper pair on the island exerts a static electric field, coupling to a molecule positioned in the waveguide, its strength depending on the spacing $d$ and the relative position of the molecule. (b) Numerical evaluation of electric field strength $\Delta\mathcal{E}$ at the position of the molecule, depending on the distance $d$, for W = $700$ nm and two different heights $\text{H}$ of the waveguide (solid lines: $H=200~\text{nm}$, dashed lines: $H=400~\text{nm}$) and three relative position of the molecule (blue, black and red) corresponding to the schematic in (a)}
\label{waveguideSchematic}	
\end{figure}
In this section we estimate the qubit-molecule coupling strength $\text{g}_\text{c}$ for a simple architecture of the waveguide-molecule and qubit system as shown schematically in Fig. S$1$~(a).
From the definition of the coupling in the main text the figure of merit for this estimate is the electric field strength exerted by a single Cooper pair of charge 2e distributed over the island on the location of the molecule. We evaluate the electric field by a full $3$D numerical simulation using COMSOL Multiphysics for three different positions of the molecule inside the waveguide as depicted in Fig. S$1$~(a). For our numerical simulation we consider that the Cooper pair is distributed over an aluminum island of dimensions $700\times 300\times 25$~nm ($\text{L}_{\text{a}}\times\text{W}_{\text{a}}\times\text{H}_{\text{a}}$, where  in Fig. S$1$~(a) we show $\text{W}_{\text{a}}$ and $\text{H}_{\text{a}}$ but not $\text{L}_{\text{a}}$). Note that these parameters correspond to the size of the CPB reported in \cite{Vion02}. For simulation purpose we apply a reference voltage on the aluminum film, and using Gauss' law we adjust the voltage such that a total charge of 2e is distributed over the island. Furthermore, we simulate the structure for a waveguide width of W = $700$~nm and two different waveguide heights, $H=200$ and $400$~nm. The result of our simulation is shown in Fig. S$1$~(b), where we plot $\Delta\mathcal{E}$ as a function of the distance $d$ between the waveguide and the qubit. For an organic molecule of dipole moment $1$D as considered in the main text, one can calculate a linear Stark shift co-efficient of $5$ MHz/(kV/m). From the simulation results in Fig. S$1$~(b) we find that for our architecture if we consider the molecule to be located near the edge of the waveguide with $\text{d} \sim 125$ nm the Cooper pair on the island can create an electric field $\Delta\mathcal{E} \sim 16$~kV/m. This can thus lead to a coupling strength between the molecule and the qubit of $\text{g}_{\text{c}} = 80$ MHz as reported in the main text. From Fig. S$1$~(b) one can see that coupling strengths as high as $110-125$ MHz can in principle be achieved in such systems. 

The question of how close to the waveguide the SC qubit can be placed ultimately depends on the detrimental effects of the light on the superconductor and vice versa. 
\begin{figure} 
	\centering
	\includegraphics[width=16cm]{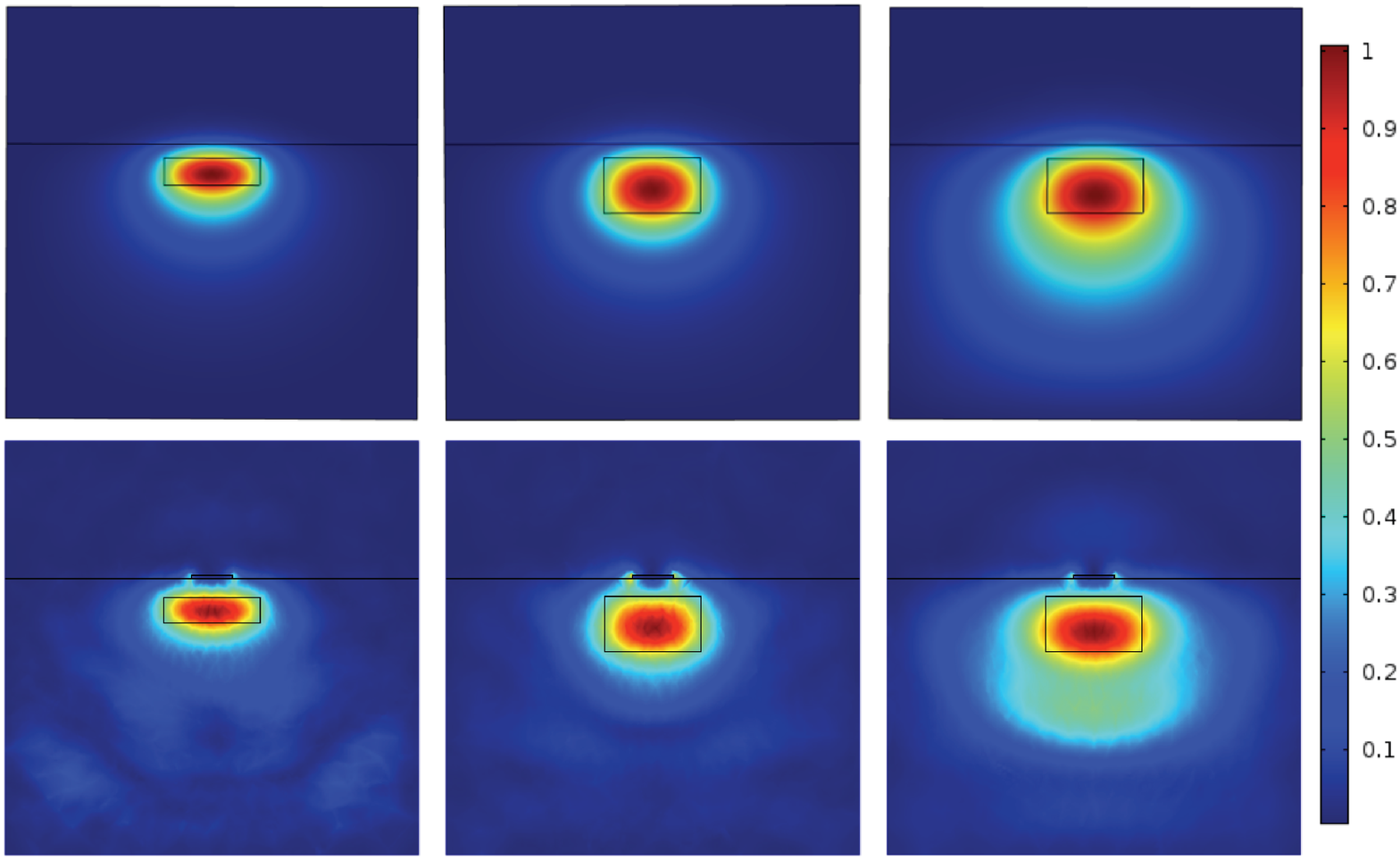}
	\caption{Waveguide mode field distributions. Top panel shows a slice of the guided mode in the absence of the qubit. This mode shape is used as a boundary condition for the full 3D simulation. The bottom panel is the result of the $3$D simulation and shows a slice through the center of the aluminum island placed at a distance of $100$~nm from the surface of the waveguide. Column-wise: Left for $\lambda=430$~nm and $H=200$~nm, middle for  $\lambda=590$~nm and $H=400$~nm and right for $\lambda=780$~nm and $H=400$~nm.}
	\label{modefield}
\end{figure}
To answer this we investigated the optical mode propagation in a polyethylene waveguide in a $\mathrm{SiO_2}$ substrate, and the influence of introducing an aluminium (metal) plate close to the waveguide. For this purpose, we performed a full $3$D numerical simulation in COMSOL Multiphysics, which solves Maxwell's equations for the electric field. We defined a cubic simulation space of dimensions $L_x\times L_y\times L_z$ where $L_x=L_y=5000$~nm and $L_z=2000$~nm, with $z$ the propagation direction. For simulation purpose we use the same setup discussed above with the aluminium island located at a distance of $100$~nm from the surface of the waveguide.

In Fig. \ref{modefield} we show the result of our numerical simulation. In the upper left panel, the waveguide has height $H=200$nm and the incoming mode is at $\lambda=430$~nm. In the middle and right cases, the waveguide has height $H=400$~nm and the incoming mode is at $\lambda=590$ and $780$~nm respectively. We evaluate the $S$-matrix for each case, to find the total amount of overlap between the exit mode and the field that has passed our simulation space. We found $S_{12}$ over $95\%$ in each case, which means that less than $5\%$ of the light was scattered or absorbed as compared to the case where there is no aluminium island. Furthermore, we evaluated the total electromagnetic power dissipation in the metal island and found values of around $1\%$ of the input power for all three cases. Thus from these results we can conclude that the presence of an aluminium island, even at close proximity of $d=100$~nm to the waveguide, has negligible impact on the device performance when using a single-photon or weak coherent pulse input: For our entanglement generation scheme discussed below we post select on the emission of a Stokes photon. If we consider an incident  single photon pulse we cannot simultaneously absorb a photon and  have a click in the detectors. Hence the influence of the aluminum will only result in a minor reduction of the success probability. With a coherent input state with $\bar n\approx 1$ there is a small probability that we can have a Raman transition simultaneous with the absorption of a photon in the aluminum. Since this only happens in $1\%$ of the cases, the qubit will be unaffected in $99\%$ of the post-selected events and the effect on the Fidelity will be limited. 

In the above discussion of the molecule-qubit coupling strength, we have considered a simplistic structure of our system with the sole purpose of explaining the rich physics of electrical coupling. However, there may be potentially several ways of improving the design of the system and engineering much better structures. This can hence lead to better/stronger coupling between the superconducting qubit and the molecule in such system. 
\section{Raman scattering Scheme} 
In this scheme we consider two molecules inside the slot-waveguide coupled to each other via optical dipole-dipole interaction. The qubit is assumed to be located near a pair of such dipole coupled molecules and is operated at the charge degeneracy point. The combination of two molecules and the qubit now represent the hybrid structure. The free energy part of the Hamiltonian of such a hybrid is similar to the single molecule case with $\mathcal{H}_{\text{m}} \rightarrow \sum_{\text{k}}\mathcal{H}^{(\text{k})}_{\text{m}}$ where the superscript $k = 1,2$ denotes the two molecules. The interaction Hamiltonian is in this case a sum of contributions from three different physical processes namely the dipole-dipole interaction $\mathcal{H}_{\text{dd}}$, the molecule-qubit interaction $\sum_{k}\mathcal{H}^{\text{I}}_{\text{mq,k}}$, and the molecule-light interaction $\sum_{k}\mathcal{H}^{\text{I}}_{\text{ml,k}}$. Following Eq. (\ref{eq1}) and Eq. (\ref{eq2}) these can be written
\bea
\label{eq12}
\mathcal{H}_{\text{I}} & = & \mathcal{H}_{\text{dd}}+\sum_{\text{k}}\mathcal{H}^{\text{I}}_{\text{ml,k}}+\sum_{k}\mathcal{H}^{\text{I}}_{\text{mq,k}}~,\nonumber\\
\mathcal{H}_{\text{dd}} & = & \hbar\text{V}(\sigma^{+}_{1}\sigma^{-}_{2}+ \sigma^{+}_{2}\sigma^{-}_{1})~,\\
\mathcal{H}^{\text{I}}_{\text{ml,k}} & = &\frac{\hbar \text{g}_{\text{m}_\text{k}}}{2}\sigma^{+}_{\text{k}}\hat{a}e^{i[kr_{\text{k}}-\omega_{\text{p}}t]}+\frac{\hbar \text{g}_{\text{m}_\text{k}}}{2}\hat{a}^{\dagger}\sigma^{-}_{\text{k}}e^{-[ikr_{\text{k}}-\omega_{\text{p}}t]}~,\\
\mathcal{H}^{\text{I}}_{\text{mq,k}} & = & \frac{\hbar\text{g}_{\text{c}_\text{k}}}{4}\eta_{\text{z}}\otimes\left(\sigma^{\text{z}}_{\text{k}}+\mathbb{I}\right)~,
\eea
where $\text{g}_{\text{m}_\text{k}}$ and $\text{g}_{\text{c}_\text{k}}$ are the coupling strength of the $\text{k}^{\text{th}}$ molecule to the incoming light and the CPB qubit respectively. The combined basis of the molecule-CPB qubit hybrid can be written as $\{|e_{1},e_{2}\rangle\otimes|\pm\rangle, |S\rangle\otimes|\pm\rangle,|A\rangle\otimes|\pm\rangle, |g_{1},g_{2}\rangle\otimes|\pm\rangle\}$. Here the index $1,2$ corresponds to the molecule and $|\pm\rangle$ are the qubit eigenstates at the energy degeneracy point while, the states $|S\rangle = \beta_{1} |e_{1}g_{2}\rangle+ \beta_{2}|g_{1}e_{2}\rangle$ and $|A\rangle = \beta'_{1} |e_{1}g_{2}\rangle - \beta'_{2}|g_{1}e_{2}\rangle$ are the eigen-states of the Hamiltonian $\mathcal{H}_{\text{dd}}$  with the co-efficients, 
\bea
\label{eq13}
\beta_{1} & = &\beta'_{2} = \sqrt{\frac{1}{2}\left(1+\frac{\delta_{0}}{\sqrt{4\text{V}^{2}+\delta^{2}_{0}}}\right)}~,\qquad \beta'_{1} = \beta_{2} = \sqrt{\frac{1}{2}\left(1-\frac{\delta_{0}}{\sqrt{4\text{V}^{2}+\delta^{2}_{0}}}\right)}~.
\eea
Here, $\delta_{0} = (\omega_{\text{m}_1}-\omega_{\text{m}_{2}})$, while the co-efficients satisfy $(\beta^{2}_{1}+\beta^{2}_{2}) = (\beta'^{2}_{1}+\beta'^{2}_{2}) = 1$, $\beta_{1}\beta_{2} = \beta'_{1}\beta'_{2} =\text{V}/\sqrt{4\text{V}^{2}+\delta^{2}_{0}}$, $(\beta_{1}^{2}-\beta_{2}^{2})= (\beta_{2}^{'2}-\beta_{1}^{'2}) = \delta_{0}/\sqrt{4\text{V}^{2}+\delta^{2}_{0}}$. We consider the incoming light pulse interacting with the molecules to be quite weak (single to few photons). Hence, two photon processes leading to excitation to the state $|e_{1},e_{2}\rangle$ can be neglected from the scattering dynamics. Thus, the basis states of the hybrid is restricted to $\{|S\rangle\otimes|\pm\rangle,|A\rangle\otimes|\pm\rangle, |g_{1},g_{2}\rangle\otimes|\pm\rangle\}$ as shown in Fig (\ref{fig2}). 
\begin{figure}[h!]                                        
 \includegraphics[height=6.5cm]{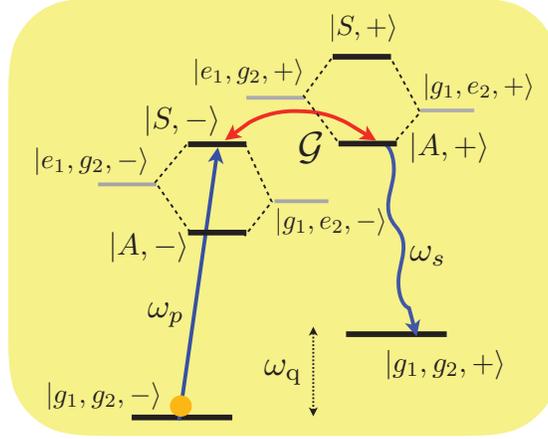}
   \caption 
   { \label{fig2} Schematic of the energy levels in the molecule-SC qubit hybrid for the Raman scattering process. The molecular levels $|e_1g_2\rangle$ and $|g_1 e_2\rangle$ are hybridized by the dipole-dipole interaction between the molecules to form the dressed states $|A\rangle$ and $|S\rangle$. The separation of these dressed states can be tuned into resonance with the qubit frequency $\omega_\text{q}$ (the energy separation between the ground states). Scattering of a photon of frequency $\omega_\text{p}$ along the transition $|g_{1},g_{2},-\rangle\rightarrow|S,-\rangle$ leads to emission of a Stokes photon $\omega_\text{s}$ along the transition $|A,+\rangle \rightarrow |g_{1},g_{2},+\rangle$ due to resonant coupling among the states $|S,-\rangle \leftrightarrow |A,+\rangle$.  }
\end{figure} 

The states $|S,-\rangle$ and $|A,-\rangle$ in Fig. \ref{fig2} have energies $\pm\mathcal{V} = \pm\frac{1}{2}\sqrt{4V^{2}+\delta^{2}_{0}}$ corresponding to an energy separation of $2\mathcal{V}$ while $|S,\pm\rangle (|A,\pm\rangle)$ are separated by the qubit transition frequency $\omega_{q}$ equal to the ground state seperation. Furthermore, the dressed states $|S,-\rangle$ and $|A,+\rangle$ have an effective coupling of $\mathcal{G} = (\text{g}_{\text{c}_1}-\text{g}_{\text{c}_2})\text{V}/\sqrt{4\text{V}^{2}+\delta^{2}_{0}}$ among them. For a dipolar interaction strength $V < \omega_\text{q}$, it is possible to vary the frequency difference among the molecules $\delta_{0}$ by external field so as to tune the energy difference among the dressed states $2\mathcal{V}$ into resonance with $\omega_\text{q}$. This resonance condition allows the exchange of energy between the qubit and the excited manifold of the molecules which thereby enables the Raman transition $|g_{1}, g_{2}, -\rangle\rightarrow |S,-\rangle\rightarrow |A,+\rangle \rightarrow |g_{1},g_{2},+\rangle$ when the hybrid interacts with an incoming photon resonant to the transition $|g_{1},g_{2},-\rangle \rightarrow |S,-\rangle$. This process is illustrated in Fig. \ref{fig2}. However, for big $\delta_{0}$ the coupling among the dressed states becomes weak and one thus needs to strike a balance between $\delta_{0}$ and $V$ when selecting two such molecule. For the hybrid structure in the Raman configuration the non-Hermitian Hamiltonian in the basis $\{|S,\pm\rangle, |A,\pm\rangle\}$ is 
\bea
\label{eq14}
\mathcal{H}^{(1)}_{\text{nh}} = \left(
\begin{array}{cccc}
 \mathcal{V}+\Delta +\omega_{q} -\frac{i \text{$\Gamma_s$}}{2} & \frac{\mathcal{G}_1}{2} & -\frac{i \Gamma_{as}}{2} & \mathcal{G} \\
 \frac{\mathcal{G}_1}{2} & \mathcal{V}+\Delta -\frac{i \text{$\Gamma_s$}}{2} & \mathcal{G} & -\frac{i \Gamma_{as}}{2} \\
 -\frac{i \Gamma_{as}}{2} & \mathcal{G} & -\mathcal{V}+\Delta +\omega_{q} -\frac{i \text{$\Gamma_a$}}{2} & \frac{\mathcal{G}_2}{2} \\
 \mathcal{G} & -\frac{i \Gamma_{as}}{2}  & \frac{\mathcal{G}_2}{2} & -\mathcal{V}+\Delta -\frac{i \text{$\Gamma_a$}}{2} \\
\end{array}
\right),
\eea
corresponding to the photon scattering along the transition $|g_{1},g_{2},-\rangle \rightarrow |S, -\rangle$, while that for scattering along the transition pathway $|g_{1},g_{2},+\rangle\rightarrow|S,+\rangle$ is
\bea
\label{eq15}
\mathcal{H}^{(2)}_{\text{nh}} = \left(
\begin{array}{cccc}
 \mathcal{V}+\Delta -\frac{i \text{$\Gamma_s$}}{2} & \frac{\mathcal{G}_1}{2} & -\frac{i\Gamma_{as}}{2} & \mathcal{G} \\
 \frac{\mathcal{G}_1}{2} & \mathcal{V}+\Delta -\omega_{q} -\frac{i \text{$\Gamma_s$}}{2} & \mathcal{G} & -\frac{i\Gamma_{as}}{2} \\
 -\frac{i \Gamma_{as}}{2} & \mathcal{G} & -\mathcal{V}+\Delta -\frac{i \text{$\Gamma _a$}}{2} & \frac{\mathcal{G}_2}{2} \\
 \mathcal{G} & -\frac{i \Gamma_{as}}{2} & \frac{\mathcal{G}_2}{2} & -\mathcal{V}+\Delta -\omega_{q} -\frac{i \text{$\Gamma_a$}}{2} \\
\end{array}
\right),
\eea
where, $\Gamma_{as} = \gamma_{c}\frac{\delta_{0}}{\sqrt{4V^{2}+\delta^{2}_{0}}}, \Gamma_{s} = \gamma+ 2\gamma_{c}\frac{V}{\sqrt{4V^{2}+\delta^{2}_{0}}}, \Gamma_{a} =  \gamma - 2\gamma_{c}\frac{V}{\sqrt{4V^{2}+\delta^{2}_{0}}}, \mathcal{G}_1= \frac{1}{2} \left(\text{g}_{\text{c}_1}+\text{g}_{\text{c}_2}+\frac{\delta _0 \left(g_{\text{c}_1}-g_{\text{c}_2}\right)}{\sqrt{4V^2+\delta _0^2}}\right)$, $\mathcal{G}_2 = \frac{1}{2} \left(\text{g}_{\text{c}_1}+\text{g}_{\text{c}_2}-\frac{\delta _0 \left(g_{\text{c}_1}-g_{\text{c}_2}\right)}{\sqrt{4V^2+\delta _0^2}}\right)$ and we have assumed the total decay rate of each emitter ($\gamma_{k = A,B} = \gamma = \gamma_{1D}+\gamma_{k,i}+\gamma_{c}$) to be equal. Here $\gamma_{k,i}$ is the intrinsic decay of the k$^{th}$ emitter and $\gamma_{c}$ is a collective decay rate. These different decay rates reflect that the two molecules can decay both to independent reservoirs giving an intrinsic decay and to a joint reservoir giving a collective decay. For the collective decay we assume that possible energy shifts due to the coupling to the collective reservoir are included in the dipole interaction $V$. Furthermore for simplicity we assume that the two molecules have the same relative phase in their interaction with the waveguide and the common reservoir. We invoke two different Hamiltonians for the two different transitions pathways, because the initial states have different energies and thus different effective detunings \cite{flo12}. From the central block of Eq. (\ref{eq13}) and (\ref{eq14}) that involves the $|S,-\rangle \rightarrow |A,+\rangle$ and $|A,+\rangle \rightarrow |S,-\rangle$ respectively, it is clear that the resonance condition for scattering along the two paths is quite different. Thus a certain choice of the resonance condition will enhanced one transition pathway while suppressing the other.

To describe the scattering dynamics we assume that the waveguide is semi-infinite and single sided. The input-output relations Eq. (\ref{eq3}) and Eq. (\ref{eq3a}), then following Ref.\cite{dirk} reduces to
\bea
\label{eq15aa}
\hat{a}_{o}(\text{z},t) = \hat{a}_{\text{in}}(\text{z}-v_{g}t) +i\sum_{\text{m}}e^{-i\omega_{\text{mm}'}(z'-z)/v_{g}}\rho_{\text{mm}'}(t)\zeta_{\text{m}\text{m}'}\hat{a}_{\text{in}}(\text{z}-v_{g}t),
\eea
with now $\Gamma_\text{em}/2 \rightarrow \Gamma_\text{em}$. To evaluate the density matrix elements $\rho_{\text{mm}'}$ appearing in the above equation, for the process $|g,-\rangle\rightarrow|S,-\rangle\rightarrow|A,+\rangle\rightarrow|g,+\rangle$ we use the master equation derived in the effective operator formalism \cite{flo12, Das16}
\bea
\label{eq15a}
\dot{\hat{\rho}} & = &: i\left[H_{eff}, \hat{\rho}\right]-\frac{1}{2}\sum_{k}\left(\mathcal{L}^{k\dagger}_{eff}\mathcal{L}^{k}_{eff}\hat{\rho}+\hat{\rho}\mathcal{L}^{k\dagger}_{eff}\mathcal{L}^{k}_{eff}\right)+\sum_{k}\mathcal{L}^{k}_{eff}\hat{\rho}\mathcal{L}^{k\dagger}_{eff} :,
\eea
where $: ........ :$ denotes normal ordering. The effective Hamiltonian is written 
\bea
\label{eq15b}
H_{eff} & = &\frac{1}{2}\left(\text{g}_{\text{m}_1}\beta'_{2}+\text{g}_{\text{m}_2}\beta'_{1}\right)^{2}\left[\left(\mathcal{H}^{(1)}_{\text{nh}}\right)^{-1}_{22} +\left(\mathcal{H}^{(1)\dagger}_{\text{nh}}\right)^{-1}_{22}\right]|1\rangle\langle 1|\hat{a}^{\dagger}\hat{a}\nonumber\\
& + & \frac{1}{2}\left(\text{g}_{\text{m}_1}\beta_{2}-\text{g}_{\text{m}_2}\beta_{1}\right)^{2}\left[\left(\mathcal{H}^{(2)}_{\text{nh}}\right)^{-1}_{33} +\left(\mathcal{H}^{(2)\dagger}_{\text{nh}}\right)^{-1}_{33}\right]|4\rangle\langle 4|\hat{a}^{\dagger}\hat{a},\nonumber\\
\eea 
where we have introduced the convention $|1\rangle = |g_1,g_2,-\rangle$, $|2\rangle = |S,-\rangle$, $|3\rangle = |A,+\rangle$ and $|4\rangle = |g_1,g_2,+\rangle$ that will be used in all further calculations. The effective Lindbald operators are
\bea
\label{eq15c}
\mathcal{L}^{k}_{eff} = \mathcal{L}^{k}\left[\left(\mathcal{H}^{(1)}_\text{nh}\right)^{-1}+\left(\mathcal{H}^{(2)}_\text{nh}\right)^{-1}\right]V_{+}.
\eea
Here $\mathcal{L}^{k}$, depending on the situation stands for $\mathcal{L}^{\gamma_{i}}_{1}, \mathcal{L}^{\gamma_{i}}_{2}$ and $\mathcal{L}^{\gamma_{1D}+\gamma_{c}}$ and 
\bea
\label{eq15d}
V_{+} &=&(\text{g}_{\text{m}_1}\beta^{'}_{2}+\text{g}_{\text{m}_2}\beta^{'}_{1})\left( |2\rangle\langle 1|+|S,+\rangle\langle 4|\right)\hat{a}e^{i\Delta t}+(\text{g}_{\text{m}_1}\beta_{2}-\text{g}_{\text{m}_2}\beta_{1})|3\rangle\langle 4|\hat{a}e^{i(\Delta-\omega_\text{q})t}\nonumber\\
&+&(\text{g}_{\text{m}_1}\beta_{2}-\text{g}_{\text{m}_2}\beta_{1})|A,-\rangle\langle 1|\hat{a}e^{i(\Delta-\mathcal{V})t},\\
\mathcal{L}^{\gamma_{i}}_{1} & = & \sqrt{{\gamma_{i}}_{1}}\beta'_{2}\left(|1\rangle\langle 2|+|4\rangle\langle S,+|\right)+\sqrt{{\gamma_{i}}_{1}}\beta_{2}\left(|4\rangle\langle 3|+|1\rangle\langle A,-|\right),\\
\mathcal{L}^{\gamma_{i}}_{2} & = & \sqrt{{\gamma_{i}}_{2}}\beta'_{1}\left(|1\rangle\langle 2|+|4\rangle\langle S,+|\right)-\sqrt{{\gamma_{i}}_{2}}\beta_{1}\left(|4\rangle\langle 3|+|1\rangle\langle A,-|\right),\\
\mathcal{L}^{(\gamma_{1D}+\gamma_{c})} & = &\sqrt{(\gamma_{1D}+\gamma_{c})}(\beta^{'}_{2}+\beta^{'}_{1}) \left(|4\rangle\langle S,+|+|1\rangle\langle 2|\right)+\sqrt{(\gamma_{1D}+\gamma_{c})}(\beta_{2}-\beta_{1}) \left(|4\rangle\langle 3|+|1\rangle\langle A,-|\right). \nonumber\\ 
\eea
For a single photon input, we find on solving Eq. (\ref{eq15a}) that we should use
\bea
\label{eq15e}
\rho_{11}(t) & = &\rho_{11}(0),\quad  \rho_{44}(t) = \rho_{44}(0),\nonumber\\
\rho_{14}(t)& = &\rho_{14}(0)e^{i\omega_\text{q}t}
\eea
when we insert it into Eq. (\ref{eq15aa}) because of the normal ordering (note that the normal ordering formalism used here merely reflect that a single photon can only be scattered once, and hence there is no evolution in the density matrix before the scattering).  

The scattering amplitude $\zeta_{\text{m = 1}\text{m}' = 4}$ is evaluated from Eq. (\ref{eq4}) by finding the relevant inverse of the non-Hermitian Hamiltonian matrix given in Eq. (\ref{eq14}) and (\ref{eq15}). We evaluate these in a moderate coupling limit $\text{g}^{2}_{\text{c}_{1,2}}/\gamma\omega_{\text{q}} < 1$ as
\bea
\label{eq23a}
\left[H^{(1)\dagger}_{nh}\right]^{-1}_{23}\left[H^{(1)}_{nh}\right]^{-1}_{32} & = & \frac{16\mathcal{G}^{2}}{(4\mathcal{G}^{2}+\Gamma_{s}\Gamma_{a}+4[\epsilon^{2}_{2}-\epsilon^{2}_{1}])^{2}+4(\Gamma_{s}[\epsilon_{1}-\epsilon_{2}]+\Gamma_{a}[\epsilon_{1}+\epsilon_{2}])^{2}},\\
\label{eq23b}
\left[H^{(2)\dagger}_{nh}\right]^{-1}_{32}\left[H^{(2)}_{nh}\right]^{-1}_{23} & = &\frac{(\frac{1}{4}[\epsilon_1+\epsilon_2] [8\Gamma_s\mathcal{G}-4\Gamma_{as} \mathcal{G}_{1}-16 \mathcal{G}(\epsilon_1+\epsilon_2)])^{2}+(\frac{1}{8}\Gamma_s[8\Gamma_s\mathcal{G}-4 \Gamma_{as}\mathcal{G}_1-16 \mathcal{G}(\epsilon_1+\epsilon_2)])^{2}}{\omega^{4}_\text{q}(\Gamma_s^2+4[\epsilon_1+\epsilon_2]^2)^2},\nonumber\\
\\
\label{eq23c}
\left[H^{(1)\dagger}_{nh}\right]^{-1}_{22}\left[H^{(1)}_{nh}\right]^{-1}_{22} & = & \frac{16(\Gamma_{a}+2[\epsilon_{2}-\epsilon_{1}])^{2}}{(4\mathcal{G}^{2}+\Gamma_{s}\Gamma_{a}+4[\epsilon^{2}_{2}-\epsilon^{2}_{1}])^{2}+4(\Gamma_{s}[\epsilon_{1}-\epsilon_{2}]+\Gamma_{a}[\epsilon_{1}+\epsilon_{2}])^{2}},\\
\label{eq23d}
\left[H^{(2)\dagger}_{nh}\right]^{-1}_{33}\left[H^{(2)}_{nh}\right]^{-1}_{33} & = & \frac{(2[\epsilon_1+\epsilon_2] [\Gamma_s-2(\epsilon_1+\epsilon_2)])^{2}+(\Gamma _s[\Gamma_s-2 (\epsilon_1+\epsilon_2])^{2}}{\omega^{2}_\text{q}(\Gamma_s^2+4[\epsilon_1+\epsilon_2]^2)},\\
\left[H^{(2)\dagger}_{nh}\right]^{-1}_{S+S+}\left[H^{(2)}_{nh}\right]^{-1}_{S+,S+} & = &\frac{4}{\Gamma^{2}_{s}+4\left(\epsilon_{1}+\epsilon_{2}\right)^{2}},
\eea
where $\epsilon_{1}(\epsilon_{2})$ is a small variations of $\Delta (\mathcal{V})$, but $\ll \omega_\text{q}/2$. The probability of Raman stokes scattering defined as $\mathcal{P}_{R} = \zeta_{14}\zeta_{41}$ can then be written as $\mathcal{P}_{R} = (\gamma_{1D}/\gamma)^{2}\wp_{R}$, where we find on using Eq. (\ref{eq23a})
\bea
\label{eq24}
\wp_{R} = \left(\frac{\delta_{0}}{\omega_\text{q}}\right)^{2}\left[\frac{4\mathcal{G}^{2}}{\Gamma^{2}_s\Gamma^{2}_a/4\gamma^{2}+4\mathcal{G}^{2}}\right].
\eea
In arriving at the above expression we have used the optimized resonance condition $\Delta = -\omega_\text{q}/2+\mathcal{G}, \mathcal{V} = \omega_\text{q}/2$ found by putting $\epsilon_{1} = \mathcal{G}$ and $\epsilon_{2} = 0$. 

In practise it is difficult to have a perfect single photon source. As such a more realistic solution is to use a weak coherent state. In the following we study scattering of an input weak light pulse represented by a coherent state $|\alpha\rangle$ interacting with the molecule. For our scheme $\text{m,m}'$ corresponds to the levels $|g_{1},g_{2},-\rangle$ and $|g_{1},g_{2},+\rangle$. We hence find for the resonant Raman scattering process, the density matrix elements for the corresponding population and coherences as
\bea
\label{eq16}
\rho_{11}(t) &=& \rho_{11}(0)\left\{\frac{\mathcal{P}_{IR}}{\mathcal{P}_{RS}+\mathcal{P}_{IR}}+ \left(\frac{\mathcal{P}_{RS}}{\mathcal{P}_{RS}+\mathcal{P}_{IR}}\right)e^{-(\mathcal{P}_{RS}+\mathcal{P}_{IR})|\alpha|^{2}t}\right\}+\rho_{44}(0)\left(\frac{\mathcal{P}_{IR}}{\mathcal{P}_{RS}+\mathcal{P}_{IR}}\right)(1-e^{-(\mathcal{P}_{RS}+\mathcal{P}_{IR})|\alpha|^{2}t}),\nonumber\\
\\
\label{eq17}
\rho_{44}(t) &=& \rho_{44}(0)\left\{\frac{\mathcal{P}_{RS}}{\mathcal{P}_{RS}+\mathcal{P}_{IR}}+ \left(\frac{\mathcal{P}_{IR}}{\mathcal{P}_{RS}+\mathcal{P}_{IR}}\right)e^{-(\mathcal{P}_{RS}+\mathcal{P}_{IR})|\alpha|^{2}t}\right\}+\rho_{11}(0)\left(\frac{\mathcal{P}_{RS}}{\mathcal{P}_{RS}+\mathcal{P}_{IR}}\right)(1-e^{-(\mathcal{P}_{RS}+\mathcal{P}_{IR})|\alpha|^{2}t}),\nonumber\\
\\
\label{eq18}
\rho_{14}(t)&=&\rho_{14}(0)e^{(i\omega _{14}-\mathcal{P} _{c}/2)|\alpha|^{2}t},
\eea
where 
\bea
\label{eq19}
\omega_{14} & = &\omega_\text{q}+(\text{g}_{\text{m}_{1}}\beta'_{2}+\text{g}_{\text{m}_{2}}\beta'_{1})^{2}\left(\left[H^{(1)\dagger}_{nh}\right]^{-1}_{22}+\left[H^{(1)}_{nh}\right]^{-1}_{22}\right)+(\text{g}_{\text{m}_{1}}\beta'_{2}-\text{g}_{\text{m}_{2}}\beta'_{1})^{2}\left(\left[H^{(2)\dagger}_{nh}\right]^{-1}_{33}+\left[H^{(2)}_{nh}\right]^{-1}_{33}\right),\\
\label{eq20}
\mathcal{P}_{RS} & = & \left[\left(\gamma_{1D}+\gamma_{c}\right)(\beta_{2}-\beta_{1})^{2}+\gamma_{i1}\beta_{2}^{2}+\gamma_{i2}\beta_{1}^{2}\right]
(\text{g}_{\text{m}_{1}}\beta'_{2}+\text{g}_{\text{m}_{2}}\beta'_{1})^{2}\left[H^{(1)\dagger}_{nh}\right]^{-1}_{23}\left[H^{(1)}_{nh}\right]^{-1}_{32},\\
\label{eq21}
\mathcal{P}_{IR} & = &\left[\left(\gamma_{1D}+\gamma_{c}\right)(\beta^{'}_{2}+\beta^{'}_{1})^{2}+\gamma_{i1}\beta_{2}^{' 2}+\gamma_{i2}\beta_{1}^{' 2}\right]
(\text{g}_{\text{m}_{1}}\beta_{2}-\text{g}_{\text{m}_{2}}\beta_{1})^{2}\left[H^{(2)\dagger}_{nh}\right]^{-1}_{32}\left[H^{(2)}_{nh}\right]^{-1}_{23},\\
\label{eq22}
\mathcal{P}_{c} & = &\mathcal{P}_{RS}+\mathcal{P}_{IR}+\mathcal{P}_{D},\\
\label{eq22a}
\mathcal{P}_{D} & = & \left[\left(\gamma_{1D}+\gamma_{c}\right)(\beta'_{2}+\beta'_{1})^{2}+\gamma_{i1}\beta^{'2}_{2}+\gamma_{i2}\beta^{'2}_{1}\right](\text{g}_{\text{m}_{1}}\beta'_{2}+\text{g}_{\text{m}_{2}}\beta'_{1})^{2}\left|\left[H^{(1)}_{nh}\right]^{-1}_{22}-\left[H^{(2)}_{nh}\right]^{-1}_{S+S+}\right|^{2}.
\eea
The probability of Raman scattering $P_{RS}$ given in Eq. (\ref{eq20}) can be further separated into two parts, one proportional to the probability of Raman scattering $\mathcal{P}_{R}$ into the waveguide while the other is proportional to the probability of Raman scattering $\mathcal{P}_{RO}$ to the outside which include processes where a photon is lost after scattering. On using Eqs. (\ref{eq23a}) and (\ref{eq23b}) in Eqs. (\ref{eq20}) and (\ref{eq21}), we find the probability of Raman Stokes scattering into the waveguide mode to be $\mathcal{P}_R = (\gamma_{1D}/\gamma)^{2}\wp_{R}$ where $\wp_{R}$ is given in Eq. (\ref{eq24}). The Raman scattering to modes other than the waveguide is found to be
\bea
\label{eq24a}
\mathcal{P}_{RO} = \left(\frac{\gamma_{1D}}{\gamma}\right)\left(\frac{\gamma_{c}}{\gamma}\right)\left(\frac{\delta_{0}}{\omega_\text{q}}\right)^{2}\left[\frac{2\mathcal{G}^{2}}{\Gamma^{2}_s\Gamma^{2}_a/4\gamma^{2}+4\mathcal{G}^{2}}\right]+\left(\frac{\gamma_{1D}}{\gamma}\right)\left(\frac{\gamma_{i}}{\gamma}\right)\left(1+\frac{2V}{\omega_\text{q}}\right)\left[\frac{2\mathcal{G}^{2}}{\Gamma^{2}_s\Gamma^{2}_a/4\gamma^{2}+4\mathcal{G}^{2}}\right],
\eea
Note that in deriving the above expression we have assumed $\gamma_{i,1} = \gamma_{i,2} = \gamma_{i}$ and will use this assumption throughout the remaining part of the supplementary. On evaluating the probability of inverse Raman scattering from state $|4\rangle$ to $|1\rangle$ in (\ref{eq21}) we find that, 
\bea
\label{eq24b}
\mathcal{P}_{IR} = \left[\left(\frac{\gamma_{1D}}{\gamma}\right)^{2}\left(\frac{\delta_{0}}{\omega_\text{q}}\right)^{2}+\left(\frac{\gamma_{1D}}{\gamma}\right)\left(\frac{\gamma_{c}}{\gamma}\right)\left(\frac{\delta_{0}}{\omega_\text{q}}\right)^{2}+\left(\frac{\gamma_{1D}}{\gamma}\right)\left(\frac{\gamma_{i}}{\gamma}\right)\left(1-\frac{2\text{V}}{\omega_\text{q}}\right)\right]\left(\frac{\mathcal{G}^{2}}{\gamma\omega_\text{q}}\right)^{2}\frac{\left(1+\Gamma_{as}\mathcal{G}_{1}/4\mathcal{G}^{2}-\Gamma_{s}/2\mathcal{G}\right)^{2}}{\Gamma^{2}_s/\gamma^{2}+4\mathcal{G}^{2}/\gamma^{2}}.\nonumber\\
\eea
After some algebra we get in the leading order, $\mathcal{P}_{IR}/\mathcal{P}_{RS} \propto \left(\mathcal{G}\gamma\right)^{2}/\omega^{4}_\text{q}$. Hence under the chosen resonance condition the Raman stokes process dominates over the inverse Raman process. This can also be understood from the above matrices in Eq. (\ref{eq14}) and (\ref{eq15}), where one finds from the inverse of the elements of the central blocks that the transition $|S,-\rangle \rightarrow |A,+\rangle$ dominates the scattering process for the above mentioned set of resonance condition. For all further use of the Raman scattering we will thus neglect $\mathcal{P}_{IR}$. Finally we evaluate the probability of light induced dephasing $\mathcal{P}_{D}$ due to elastic Rayleigh scattering,
\bea
\label{eq24c}
\mathcal{P}_{D} & = &\left(\frac{\gamma_{1D}}{\gamma}\right)\left[\left(\frac{\gamma_{i}}{\gamma}\right)\left(1+\frac{2\text{V}}{\omega_{\text{q}}}\right)+\left(\frac{\gamma_{1D}+\gamma_{c}}{\gamma}\right)\left(1+\frac{2\text{V}}{\omega_{\text{q}}}\right)^{2}\right] \frac{64 \mathcal{G}^4-16\Gamma_a\mathcal{G}^2 (4 \mathcal{G}-\Gamma_a)+4 \mathcal{G}^2-8\Gamma_s^2\Gamma_a\mathcal{G}}{\left(4\mathcal{G}^2+\Gamma_s^2\right)\left(\Gamma_{s}^2\Gamma_{s}^{2}/4\gamma^{2}+\mathcal{G}^2\right)}.
\eea
 
To get the plot of Fig. 2 in the main text, we express the Raman probability as a function of the ratio of the dipolar coupling between the molecules and the SC qubit transition frequency $\text{V}/\omega_{\text{q}}$, and the couplings of the molecule to the SC qubits $\left(\text{g}_{\text{c}_1}-\text{g}_{\text{c}_2}\right)$. By using the resonance conditions for optimization of the the Raman process we can write $\delta^{2}_{0} = \omega^{2}_\text{q}-4\text{V}^{2}$. Substituting this into the expression for $\mathcal{P}_{R}$ in Eq. (\ref{eq24}) we get
\bea
\label{eq41a}
\mathcal{P}_{R}\left(\gamma/\gamma_{1D}\right)^{2} = \frac{16\left(1-4y^2\right)y^{2}x^{2}}{16y^2 x^2+\left(1-4y^2\left(\frac{\gamma_{c}}{\gamma}\right)^{2}\right)^2},
\eea
where $y = \left(\text{g}_{\text{c}_1}-\text{g}_{\text{c}_2}\right)/\gamma$ and $x = \text{V}/\omega_{\text{q}}$. 
\section{Entanglement generation between a Hybrid and a photon} 
\begin{figure}[h!]                                        
  \includegraphics[height=6.5cm]{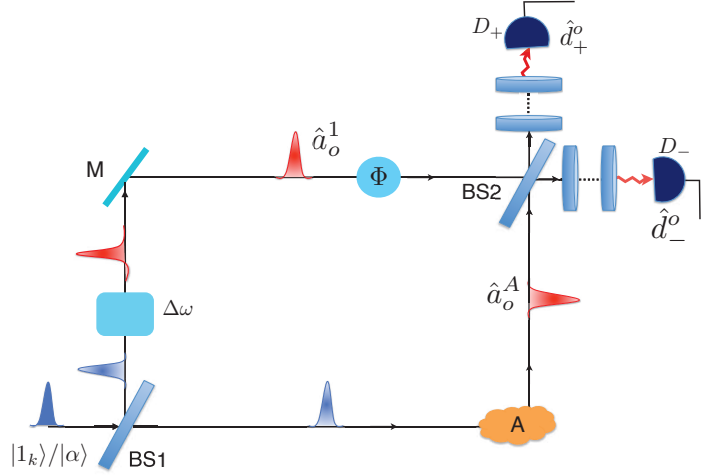}
   \caption
   { \label{fig3} 
Interferometric scheme to create entanglement between photons and the hybrid $A$. The incoming photon pulse after passing through beam splitter BS$1$ is spatially separated and travels along two arms of the interferometer. One of the component passes through a frequency modulator $\Delta\omega = \omega_{\text{q}}$ and an optical element that introduces an addition phase $\Phi$ to interfere at beam splitter BS$2$ with the other component which undergoes a Raman scattering. The two detectors $D_{\pm}$ then does a joint measurement of the photon in a basis determined by the beamsplitter and the phase $\Phi$ to setup a CHSH inequality violation. Violation of the inequality proves entanglement between the photons and the hybrid emitter A.}
\end{figure}

We first investigate entanglement between a stationary qubit and a photonic qubit by entangling the hybrid and a single photon in an interferometric setup via post-selection of scattering events. A similar scheme has been shown to achieve a perfect gate no matter how bad the light-matter coupling is \cite{chang}. As we will show in the following, we can achieve perfect operation similar to what was reported in Ref. \cite{chang}. The schematic of the entangling mechanism is depicted in Fig. \ref{fig3} . The hybrid is considered to be in the Raman configuration as shown in Fig. \ref{fig2} and forms one arm of the interferometer. Physically the entanglement creation can be understood as follows. An incoming single photon pulse $\hat{a}_{\text{in}}$, after passing through the beam splitter BS$1$ is spatially separated into two components $\hat{a}_{1}$ and $\hat{a}_{2}$. The $\hat{a}_{2}$ component is scattered from the hybrid $\text{A}$ resulting in a scattered photon $\hat{a}^{A}_{\text{o}}$. The other component $\hat{a}_{1}$, travels along the other arm of the interferometer, and gets frequency modulated by the modulator with frequency $\Delta\omega = \omega_\text{q}$ and also acquires a phase $\phi$, while passing through the phase shifter to become $\hat{a}^{1}_{o}$. The two output components $\hat{a}^{A}_{o}$ and $\hat{a}^{1}_{o}$ then interfere at the beam splitter BS$2$ coherently to form the detector mode operators $\hat{d}^{o}_{\pm}$. The photons at the two output ports of BS$2$ are collected by the single photon detectors $\text{D}_{\pm}$. If the hybrid is initialized in the state $|g,-\rangle = |g_1,g_2,-\rangle = |1\rangle$, then post-selecting the events where there is scattering, as we shall show below, leads to an entangled state of matter qubit and photonic qubit wriiten as,
\bea
\label{eq25}
|\Psi^{+}_\text{s}\rangle = \frac{1}{\sqrt{2}}\left(|U_{k}\rangle|1\rangle + |L_{k}\rangle|4\rangle\right),
\eea
where, $|U_{k}\rangle$ and $|L_{k}\rangle$ represent respectively a photon reflected from BS$1$ and a photon which has undergone Raman scattering i.e. a photon in the upper and lower arm. For a balanced interferometer, a click on the single photon detectors after the phase $\Phi$ have been applied then project the hybrid into a superposition of the lower states $|\Psi^{\pm}\rangle = \frac{1}{\sqrt{2}}(|1\rangle\pm e^{i\Phi} |4\rangle)$, depending on which of the detectors $D_{\pm}$ clicks. The post selected dynamics conditioned on the detection of a frequency shifted single photon is thus completely equivalent to the  dynamics of a maximally entangled state and allow e.g., the violation of Bell's inequality. 
 
We next mathematically treat the interferometric creation of entanglement and verify it via a Bell inequality violation corresponding to the entangled state $|\Psi^{+}_{\text{s}}\rangle$. For hybrid A initially in the state $|\Psi_{\text{ini}}\rangle = |g,-\rangle = |1\rangle$, we find the amplitude to be in some state $|j\rangle$ conditioned on detection of a single photon at the detectors $D_{\pm}$ to be given by
\bea
\label{eq26}
\mathcal{C}_j = \langle j,\O|\hat{d}^{o}_{+}(t)U(t)\hat{d}^{\text{in}\dagger}|\Psi_{\text{ini}},\O\rangle
\eea
Here, the input and output field mode operators $\hat{d}^{\text{in}}$ and $\hat{d}^{o}_{\pm}$ respectively are defined by, 
\bea
\label{eq27}
\hat{d}_{\pm}^{o}(t) & = & \frac{1}{\sqrt{2}}\sqrt{\eta}(e^{i\Phi}\hat{a}^{1}_{o}(t)\pm\hat{a}^{A}_{o}(t)) +\hat{\mathcal{F}}\\
\label{eq27a}
\hat{d}^{\text{in}} & = & (e^{i\omega_{\text{q}}t}\chi\hat{a}_{1} +\sqrt{1-\chi^{2}}\hat{a}_{2})
\eea
where in writing Eq. (\ref{eq27a}) we assumed the beam splitter BS$1$ to be asymmetric with $\chi$ as the asymmetric co-efficient, $\hat{\mathcal{F}}$ is the noise operator and $\eta$ is the photo detection efficiency of the single photon detectors. The exponential factor in (\ref{eq27a}) describes the effect of the modulator, and 
\bea
\label{eq28}
\hat{a}^{A}_{o}(t) & = & \hat{a}_{2}+ie^{-i\omega_\text{q}(z_{A}-z)/v_{g}}\zeta_{41}\rho_{14}(t)\hat{a}_{2},\nonumber\\
\hat{a}^{1}_{o}(t) & = & \hat{a}_1
\eea
Furthermore, we have assumed a semi-infinite single sided waveguide and have used the input-output relation of Eq. (\ref{eq15aa}) with $\zeta_{41}$ defined in Eq. (\ref{eq4}) while $\rho_{14}(t)$ for a single photon input is given by Eq. (\ref{eq15e}).

From Eq. (\ref{eq26}) we can write down the corresponding normalized density matrix elements as,
\bea
\label{eq28a}
\rho^{\pm}_{\text{s}_{ij}} & = & \frac{\text{Tr}\left(|i\rangle\langle j|\hat{d}^{o}_{\pm}(t)U(t)d^{\text{in}\dagger}|\Psi_{\text{ini}},\O\rangle\langle\Psi_{\text{ini}},\O|d^{\text{in}}U^{\dagger}(t)\hat{d}^{o\dagger}_{\pm}(t)\right)}{\text{Tr}(\rho^{\pm}_{\text{s}})},\nonumber\\
& = &\frac{\langle\Psi_{\text{ini}},\varnothing|\hat{d}^{\text{in}}U^{\dagger}(t)\hat{d}^{o\dagger}_{\pm}(t)U(t)U^\dagger(t)|i\rangle\langle j|U(t)U^{\dagger}(t)\hat{d}^{o}_{\pm}(t)U(t)\hat{d}^{\text{in}\dagger}|\Psi_{\text{ini}},\varnothing\rangle}{\text{Tr}(\rho^{\pm}_{\text{s}})},\nonumber\\
& = & \frac{\langle\Psi_{\text{ini}},\O|\hat{d}^{\text{in}}\hat{d}^{o\dagger}_{H,\pm}(t)\rho^{H}_{ij}(t)\hat{d}^{0}_{H,\pm}(t) \hat{d}^{in,\dagger}|\Psi_\text{ini},\O\rangle}{\text{Tr}(\rho^{\pm}_{\text{s}})}.
\eea
Here, $\text{Tr}_{f}$ is the trace over all the field modes and the superscript/subscript $H$ stand for Heisenberg picture. For all later reference we will drop this subscript/superscript with the underlying assumption that all the operator evolution is in the Heisenberg picture. Note that we here conditioned on a single detection at a time $t$. Since for now we only consider a single incident photon at most a single photon can come out and this provide a complete characterization of the output.  
On evaluating Eq. (\ref{eq28a}) we find the components of $\rho^{\pm}_{\text{s}}$ to be
\bea
\label{eq29}
\rho^{\pm}_{\text{s},11} = \frac{1}{2},\qquad \rho^{\pm}_{\text{s},44} = \frac{1}{2}, \qquad\rho^{\pm}_{\text{s},14} = \mp\frac{1}{2}ie^{-i\omega_\text{q}T}e^{i\Phi},
\eea 
where we assume the interferometer to be balanced such that equal intensities arrives at BS$2$ from both arms (see below) and all other phases arising from the arms of the interferometer have been absorbed. Note that after the detection at time $t$, the density matrix should be propagated to the final time $T$. Combining this with the phase evolution appearing in (\ref{eq27a}) leads to a total relative phase of $e^{-i\omega_{\text{q}}T}$ as seen in the above equation. We will omit this phase for all further calculations as it merely reflect the fact that the density matrix is not in the interaction picture with respect to $\mathcal{H}_{0}$.

To check the quantum correlation among the hybrid and the photon and thereby the entanglement of the state $|\Psi_{\text{ent}}\rangle$ we next consider a Bell-CHSH inequality  \cite{CHSH1,CHSH2} violation involving single photon detection at the detectors $\text{D}_{\pm}$. Projecting the density matrix $\rho^{\pm}_{\text{s}}$ on the state $|\Psi^{\pm}_{m}\rangle = \frac{1}{\sqrt{2}}\left(|1\rangle \pm e^{i\Psi}|4\rangle\right)$ in the measurement basis characterized by the angle $\Psi$ we find the joint probability of qubit detection and photodetection at the detectors $\text{D}_{\pm}$ given by $P_{\pm\pm}$ to be
\bea
\label{eq31}
P_{++} & = & P_{--} = \frac{1}{2}\cdot\frac{1}{2}\left[1+\sin\left( \Psi-\Phi\right)\right],\nonumber\\
P_{+-} & = & P_{-+} = \frac{1}{2}\cdot\frac{1}{2}\left[1-\sin\left( \Psi-\Phi\right)\right].
\eea
Here, the first and second subscript of $P$ stands for the photon detection by a respective detector and projection of the hybrid to either of the states $|\Psi^{\pm}_{\text{s}}\rangle$. The measurement outcome for certain choice of phases can then be written in spirit of the Bell inequality as, 
\bea
\label{eq32}
E(\Psi_{\text{a}},\theta_{\text{b}}) & = & \frac{P_{++}+P_{--}-P_{+-}-P_{-+}}{P_{++}+P_{--}+P_{+-}+P_{-+}} = \sin\left( \Psi_\text{a}-\Phi_\text{b}\right). 
\eea
The Bell inequality violation parameter can then be defined as
\bea
\label{eq33}
S = E(\Psi_{\text{a}},\theta_{\text{b}})-E(\Psi_{\text{a}},\theta_{\text{b}'})+E(\Psi_{\text{a}'},\theta_{\text{b}})+E(\Psi_{\text{a}'},\theta_{\text{b}}'), 
\eea
and we get a maximal violation $S  = 2\sqrt{2}$ for the following set of phase angles $\{\Psi_{\text{a}},\Psi_{\text{a}'},\Phi_{\text{b}},\Phi_{\text{b}'}\} = \{\pi/4, 3\pi/4,0,\pi/2\}$. 
The corresponding success probability is given by
\bea
\label{eq33a}
P^{(1)}_{\text{suc}} & = &\langle\Psi_{\text{ini}},\O|\hat{d}^{\text{in}}\hat{d}^{o\dagger}_{\pm}(t)\hat{d}^{o}_{\pm}(t)\hat{d}^{\text{in}\dagger}|\Psi_{\text{ini}},\O\rangle\nonumber\\
& = &2\eta\zeta_{41}\zeta^{\dagger}_{41}(1-\chi^{2})
\eea 
where, $\zeta_{41}\zeta^{\dagger}_{41} =\left(\sqrt{\gamma^{43}_{1D}}(H^{(2)}_{nh})^{-1}_{32}\sqrt{\gamma^{21}_{1D}}\right)\left(\sqrt{\gamma^{12}_{1D}}(H^{(2)\dagger}_{nh})^{-1}_{23}\sqrt{\gamma^{34}_{1D}}\right)$. In writing the above expression for the success probability we add contribution from both the ± detectors as they both give the desired outcome. Furthermore we have used that the interferometer is balanced such that equal intensities are incident on BS$2$ from the two arms of the interferometers. This is achieved by $\zeta_{41}\zeta^\dagger_{41}(1-\chi^2) = \chi^{2}$. On using the resonance conditions along with Eq. (\ref{eq23a}) in Eq. (\ref{eq33a}) we get 
\bea
\label{eq33b}
P^{(1)}_{\text{suc}} =2\eta\left(\frac{\mathcal{P}_{R}}{1+\mathcal{P}_{R}}\right),
\eea
where $\mathcal{P}_{R}$ is given in Eq. (\ref{eq24}).

If the incoming photon pulse $\hat{a}_{\text{in}}$ is assumed to be in a coherent state $|\alpha\rangle$ then Eq. (\ref{eq28a}) becomes, 
\bea
\label{eq34}
\rho^{(\pm)}_{\text{s}_{ij}} = \frac{\langle\Psi_{\text{ini}},\alpha|\hat{d}^{o\dagger}_{\pm}(t)\rho^{\pm}_{ij}(t)\hat{d}^{o}_{\pm}(t)|\Psi_{\text{ini}},\alpha\rangle}{\text{Tr}(\rho^{\pm}_{\text{s}})},
\eea
In Eq. (\ref{eq26}) we conditioned on having a click at a certain time $t$, represented by the operators $d^o_\pm$. Experimentally one would however, only consider the first click which arrive at the detector. This makes no difference above where only a single photon is involved in the process. With an incident coherent state a more correct description would be to include in Eq. (\ref{eq34}) the requirement that there is no photon detected before the time $t$.  Since we mainly consider the limit of low $(\gamma_{1D}/\gamma)$, the probability of having two detection events in the time interval is negligible and the simple description in Eq. (\ref{eq34}) is sufficient. Following the procedure discussed in detail for the single photon input pulse, and allowing for the density matrix $\rho^{(\pm)}_{\text{s}_{ij}}$ to evolve following Eqs. (\ref{eq16}) - (\ref{eq18}) we arrive at a CHSH measurement outcome of 
\bea
\label{eq35}
E(\Psi_{\text{a}},\theta_{\text{b}}) & = & \frac{P_{++}+P_{--}-P_{+-}-P_{-+}}{P_{++}+P_{--}+P_{+-}+P_{-+}} = \left[\frac{2\chi\sqrt{1-\chi^{2}}~e^{-(\mathcal{P}_{R}+\mathcal{P}_{RO})|\alpha|^{2}(1-\chi^{2})t}e^{-(\mathcal{P}_{R}+\mathcal{P}_{RO}+\mathcal{P}_{D})|\alpha|^{2}(1-\chi^{2})(T-t)/2}}{\left(\chi^{2}+\mathcal{P}_{R}(1-\chi^{2})e^{-(\mathcal{P}_{R}+\mathcal{P}_{RO})|\alpha|^{2}(1-\chi^{2})t}\right)}\right]\nonumber\\
&&\qquad\qquad\qquad\qquad\qquad\qquad\times\sin\left( \Psi_\text{a}-\Phi_\text{b}\right),
\eea 
where we allow for decoherence of the hybrids from the detection at time $t$ to the final time of the pulse $T$. In such a situation one needs to evaluate the average of the CHSH measurement in the form
\bea
\label{eq35a}
\bar{E}(\Psi_{\text{a}},\theta_{\text{b}}) = \frac{\int^{T}_{0}E(\Psi_{\text{a}},\theta_{\text{b}})p(t)dt}{P^{(c)}_{\text{suc}}},
\eea
where $p(t) =\eta|\alpha|^{2}\left(\chi^{2}+\mathcal{P}_{R}(1-\chi^{2})e^{-(\mathcal{P}_{R}+\mathcal{P}_{RO})|\alpha|^{2}(1-\chi^{2})t}\right)$ and the success probability is given by $P^{(c)}_{\text{suc}} = \int^{T}_{0}p(t)dt$. Substituting this measurement outcome into the Bell inequality of Eq. (\ref{eq33}) then gives us the violation parameter as 
\bea
\label{eq36}
S & = & 2\sqrt{2}\times 4\left(\frac{\mathcal{P}_{R}+\mathcal{P}_{RO}}{\mathcal{P}_{R}+\mathcal{P}_{RO}-\mathcal{P}_{D}}\right)\left[\frac{\mathcal{P}_{R}e^{-\frac{\bar{n}}{2}(\mathcal{P}_{R}+\mathcal{P}_{RO}+\mathcal{P}_{D})(1-\chi^{2})}\left(1-e^{-\frac{\bar{n}}{2}(\mathcal{P}_{R}+\mathcal{P}_{RO}-\mathcal{P}_{D})(1-\chi^{2})}\right)}{(\mathcal{P}_{R}+\mathcal{P}_{RO})\bar{n}\chi^{2}+\mathcal{P}_{R}\left(1-e^{-(\mathcal{P}_{R}+\mathcal{P}_{RO})\bar{n}(1-\chi^{2})}\right)}\right]\nonumber\\
& = &2\sqrt{2}\times 4\left(\frac{\mathcal{P}_{R}+\mathcal{P}_{RO}}{\mathcal{P}_{R}+\mathcal{P}_{RO}-\mathcal{P}_{D}}\right)\left[\frac{\mathcal{P}_{R}e^{-\frac{\bar{n}}{2}(\mathcal{P}_{R}+\mathcal{P}_{RO}+\mathcal{P}_{D})/(1+\mathcal{P}_{R})}\left(1-e^{-\frac{\bar{n}}{2}(\mathcal{P}_{R}+\mathcal{P}_{RO}-\mathcal{P}_{D})/(1+\mathcal{P}_{R})}\right)}{(\mathcal{P}_{R}+\mathcal{P}_{RO})\bar{n}\left(\frac{\mathcal{P}_{R}}{1+\mathcal{P}_{R}}\right)+\mathcal{P}_{R}\left(1-e^{-(\mathcal{P}_{R}+\mathcal{P}_{RO})\bar{n}/(1+\mathcal{P}_{R})}\right)}\right]
\eea
where $\bar{n}$ is the mean number of photons involved in the scattering process. The corresponding success probability is given by
\bea
\label{eq36a}
P^{(c)}_{\text{suc}} & = &\eta\left\{\bar{n}\chi^{2}+\frac{\mathcal{P}_{R}}{\mathcal{P}_{R}+\mathcal{P}_{RO}}(1-e^{-(\mathcal{P}_{R}+\mathcal{P}_{RO})\bar{n}(1-\chi^{2})})\right\}\nonumber\\
& = &\eta\left\{\bar{n}\left(\frac{\mathcal{P}_{R}}{1+\mathcal{P}_{R}} \right)+\frac{\mathcal{P}_{R}}{\mathcal{P}_{R}+\mathcal{P}_{RO}}(1-e^{-(\mathcal{P}_{R}+\mathcal{P}_{RO})\frac{\bar{n}}{1+\mathcal{P}_{R}}})\right\}
\eea 
In Fig. 3(c) of the main text we plot the Eqs. (\ref{eq36}) and (\ref{eq36a}). 
\section{Entanglement generation between two hybrids} 
To entangled two hybrids we consider a similar interferometric setup to that shown schematically in Fig. \ref{fig3}(a) but now with hybrids A and B in both the arms of the interferometer and BS$1$ is a $50-50$ beam splitter as shown in Fig. 3(a) of the main text. The physics behind the generation of entanglement has been detailed in the main text. 

First we consider an incident single photon state and evaluate the fidelity and success probabililty of the entangled state. The hybrids are initially prepared in the state $|\Psi_{\text{ini}}\rangle = |g,-\rangle_{\text{A}}\otimes|g,-\rangle_{\text{B}} = |1\rangle_{\text{A}}|1\rangle_{\text{B}} $. Due to Raman scattering of a single photon the hybrids evolves to the entangled state $|\Psi_{\pm}\rangle$, conditioned on the detection of a photon in either of the detectors $\text{D}_{\pm}$. The fidelity $F = \langle\Psi_{\pm}|\rho^{\pm}_{\text{AB}}|\Psi_{\pm}\rangle$ of the state $|\Psi_{\pm}\rangle$ can be evaluated by finding the time evolved density matrix components,
\bea
\label{eq37}
\rho^{\pm}_{\text{AB}_{ij}} = \frac{\langle\Psi_{\text{ini}},\O|\hat{d}^{in}\hat{d}^{o\dagger}_{\pm}(t)\rho^{\pm}_{ij}(t)\hat{d}^{o}_{\pm}(t)\hat{d}^{in\dagger}|\Psi_{\text{ini}}, \O\rangle}{\text{Tr}(\rho^{\pm}_{\text{AB}})},
\eea
where now $\rho_{ij} = |i_{AB}\rangle\langle j_{AB}|$ and the input and output field mode operators are defined respectively by
\bea
\label{eq38}
\hat{d}_{\pm}^{o}(t) & = &\frac{1}{\sqrt{2}}\sqrt{\eta}(\hat{a}^{\text{A}}_{o}(t)\pm\hat{a}^{\text{B}}_{o}(t))+\mathcal{F},\nonumber\\
\hat{d}^{\text{in}} & = & \frac{1}{\sqrt{2}}(\hat{a}^{\text{A}}_{1}+\hat{a}^{\text{B}}_{1}).
\eea
and the The input-output relation of Eq. (\ref{eq15aa}) gives
\bea
\label{eq39}
\hat{a}^{j}_{o}(t) & = &\hat{a}^{j}_{1}+i e^{-i\omega_{\text{q}}(z_{j}-z)/v_{g}}\zeta^{j}_{41}\rho^{j}_{14}(t)\hat{a}^{j}_{1},
\eea
where, $\zeta_{41}$ can be evaluated following Eq. (\ref{eq4}). Substituting Eq. (\ref{eq38}) in Eq. (\ref{eq37}) and on using Eq. (\ref{eq39}) and considering identical characteristics for the hybrid we find 
\bea
\label{eq40}
\rho^{\pm}_{\text{AB}_{11}} = \frac{1}{2},\qquad \rho^{\pm}_{\text{AB}_{44}} = \frac{1}{2}, \qquad \rho^{\pm}_{\text{AB}_{41,14}} = \pm \frac{1}{2}.
\eea
For detection at $D_{-}$- the quality of the entangled state is characterized by the fidelity $F = \langle\Psi_{-}|\rho^{-}_{AB}(t)|\Psi_{-}\rangle$ which attains the ideal value of $F = 1$. The corresponding success probability is given by
\bea
\label{eq40}
P^{(1)}_{\text{suc}} & = & \langle\Psi_{\text{ini}},\O|\hat{d}^{\text{in}}\hat{d}^{o\dagger}_{-}(t)\hat{d}^{o}_{-}(t)\hat{d}^{\text{in}\dagger}|\Psi_{\text{ini}},\O\rangle\nonumber\\
& = &\frac{1}{2}\eta\zeta_{41}\zeta^{\dagger}_{41} = \frac{1}{2}\eta\left(\sqrt{\gamma^{43}_{1D}}(H^{(2)}_{nh})^{-1}_{32}\sqrt{\gamma^{21}_{1D}}\right)\left(\sqrt{\gamma^{12}_{1D}}(H^{(2)\dagger}_{nh})^{-1}_{23}\sqrt{\gamma^{34}_{1D}}\right),
\eea 
which on adding contribution from both the detectors and using the resonance conditions along with Eq. (\ref{eq23a}) gives us 
\bea
\label{eq41}
P^{(1)}_{\text{suc}} = \eta\mathcal{P}_{R},
\eea
where $\mathcal{P}_{R}$ is given in Eq. (\ref{eq24}). In writing the above expression for the success probability we add the contributions from both the $\pm$ detectors as they both give the desired outcome.   

\begin{figure}[h!]                                        
   \begin{center}
  \includegraphics[height=6cm]{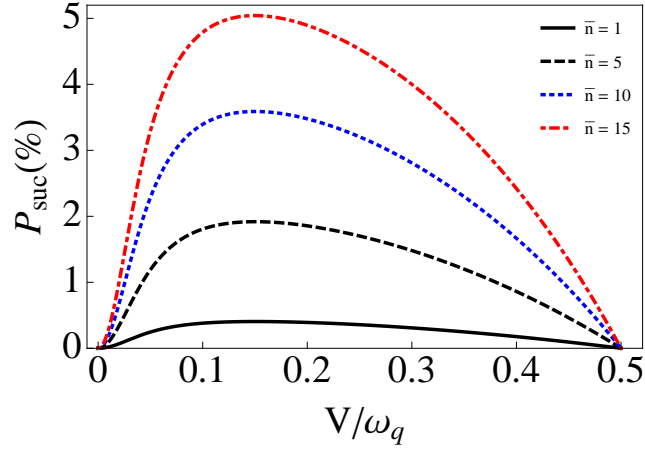}
    \end{center}
   \caption[example] 
   { \label{fig4} $P_{\text{suc}}$ as a function of $\text{V}/\omega_{\text{q}}$ for different values of the mean photon number in the incoming photon pulse. We have assumed $\gamma_{c}/\gamma = 0.45, \gamma^{i}/\gamma = 0.45$, $\gamma_{1D}/\gamma = 0.1$ and $\left(\text{g}_{\text{c}_1}-\text{g}_{\text{c}_2}\right)/\gamma = 4$.} 
\end{figure} 

If the incoming photon pulse is assumed to be in a coherent state $|\alpha\rangle$, the above treatment for evaluating the fidelity and success probability remains valid with some modifications. The components of the density matrix for the state $|\Psi_{\pm}\rangle$ now becomes,
\bea
\label{eq42}
\rho^{\pm}_{\text{AB}_{ij}} = \frac{\langle\Psi_{\text{ini}},\alpha|\hat{d}^{o\dagger}_{\pm}(t)\rho^{\pm}_{ij}(t)\hat{d}^{o}_{\pm}(t)|\Psi_{\text{ini}},\alpha\rangle}{\text{Tr}(\rho^{\pm}_{\text{AB}})},
\eea
while one evaluates now an average fidelity defined by $F = \int^{T}_{0}\langle \Psi_{-}|\rho^{-}_{\text{AB}}(t)|\Psi_{-}\rangle~p(t)dt/P^{(c)}_{\text{suc}}$. This is done to accommodate for the fact that the pulse duration $T$ is greater than the time for the click and hence there can be dephasing of the hybrid as it evolve freely during the rest of the pulse duration. The corresponding success probability is given by $P^{(c)}_{\text{suc}} = \int^{T}_{0}p(t)dt$, where $p(t) = \langle\Psi_{\text{ini}},\alpha|\hat{d}^{o\dagger}_{\pm}(t)\hat{d}^{o}_{\pm}(t)|\Psi_{\text{ini}},\alpha\rangle$. For the entangled stated $|\Psi_{-}\rangle$ we find on using the input-output relation under the resonance condition the average fidelity to be, 
\bea
\label{eq43}
F = \frac{1}{2}e^{-(\mathcal{P}_{R}+\mathcal{P}_{RO})\frac{\bar{n}}{2}}\left[1+\left(\frac{\mathcal{P}_{R}+\mathcal{P}_{RO}}{\mathcal{P}_{R}+\mathcal{P}_{RO}-\mathcal{P}_{D}}\right)\frac{\left(e^{-\mathcal{P}_{D}\frac{\bar{n}}{2}}-e^{-(\mathcal{P}_{R}+\mathcal{P}_{RO})\frac{\bar{n}}{2}}\right)}{1-e^{-(\mathcal{P}_{R}+\mathcal{P}_{RO})\frac{\bar{n}}{2}}} \right]
\eea 
with the success probability 
\bea
\label{eq44}
P^{(c)}_{\text{suc}}  & = &\frac{2P^{(1)}_{\text{suc}}}{\mathcal{P}_{R}+\mathcal{P}_{RO}}\left(1-e^{-(\mathcal{P}_{R}+\mathcal{P}_{RO})\frac{\bar{n}}{2}}\right).
\eea
To the lowest order we find the error due to dephasing of the hybrid in $\bar{F}$ to be 
\bea
\label{eq45}
F = 1-\left(\mathcal{P}_{R}+\mathcal{P}_{RO}+\frac{\mathcal{P}_{D}}{4}\right)\frac{\bar{n}}{2}
\eea
On substituting Eq. (\ref{eq41}) for $P^{(1)}_{\text{suc}}$ in the above equation we get the success probability as
\bea
\label{eq46}
P^{(c)}_{\text{suc}} = \frac{2\eta\mathcal{P}_{R}}{\left(\mathcal{P}_{R}+\mathcal{P}_{RO}\right)}\left\{1-e^{-\frac{\bar{n}}{2}\left(\mathcal{P}_{R}+\mathcal{P}_{RO}\right)}\right\}.
\eea
To the lowest order in expansion of the exponential we find $P_{\text{suc}} = \bar{n}P^{(1)}_{\text{suc}}$. We plot $P^{(c)}_{\text{suc}}$  as a function of the ratio between the dipole coupling between the molecules and the SC qubit transition energy $(V/\omega_\text{q})$ in Fig. \ref{fig4} for different values of the mean photon number. We find that the $P^{(c)}_\text{suc}$ increases significantly with the mean number of photons. As $V/\omega_\text{q} \rightarrow 1/2$, the antisymmetric state $|A\rangle$ in Fig. (S2) becomes decoupled from the dynamics of the rest of the system and hence the probability of Raman scattering vanishes $\mathcal{P}_{R} \longrightarrow 0$ which thereby leads to vanishing success probability.

Note that it is quite straightforward to include the natural dephasing of the superconducting qubit in the current framework. This is achieved by simply replacing $\frac{1}{2}\mathcal{P}_{c}|\alpha|^{2}t$ in Eq. (\ref{eq18}) with $\frac{1}{2}\mathcal{P}_{c}|\alpha|^{2}t+(t/T_{2})^{2}$, where $T_{2}$ is the coherence time of the qubit. Note that here we exploit that low frequency noise typically give rise to Gaussian decay of the coherence as observed in \cite {Wall05}. In the expression for fidelity in Eq. (\ref{eq43}) then the second term in the square bracket includes the effect of the qubit dephasing with $\mathcal{P}_{D}\frac{\bar{n}}{2}$ being modified to $\mathcal{P}_{D}\frac{\bar{n}}{2}+2(t/T_{2})^{2}$. Thus from Eq. (\ref{eq45}) we find in the lowest order, an additional reduction of the fidelity by at most an amount of $(t/T_{2})^{2}$. In reality, however, the dephasing sets in only after the detector clicks, and prepares a coherent superposition. One therefore needs to average the dephasing over the detection time. We find that in the lowest order the actual reduction of the fidelity due to qubit dephasing is only $\frac{1}{3}(T/T_{2})^{2}$ but to be conservative we use $(T/T_{2})^{2}$ in the main text.   


\begin{thebibliography}{999}

\bibitem{Scho08}
R. J. Schoelkopf and S. M. Girvin, Nature {\bf 451}, 664 (2008).

\bibitem{Dev13}
M. H. Devoret and R. J. Schoelkopf, Science {\bf 339}, 1169 (2013).

\bibitem{Martinis15}
J. Kelly, R. Barends, A. G. Fowler, A. Megrant, E. Jeffrey, T. C. White, D. Sank, J. Y. Mutus, B. Campbell, Yu Chen, Z. Chen, B. Chiaro, A. Dunsworth, I.-C. Hoi, C. Neill, P. J. J. O'Malley, C. Quintana, P. Roushan, A. Vainsencher, J. Wenner, A. N. Cleland and J. M. Martinis, Nature {\bf 519}, 66 (2015).

\bibitem{Gam15}
A.D. C—rcoles, E. Magesan, S. J. Srinivasan, A. W. Cross,	 M. Steffen, J. M. Gambetta and J. M. Chow,
Nature Communications {\bf 6},  Article number: 6979 (2015).

\bibitem{Kim08}
H. J. Kimble, Nature {\bf 453}, 1023 (2008).

\bibitem{Briegel98}
H.-J. Briegel, W. D\"ur, J. I. Cirac, and P. Zoller
Phys. Rev. Lett. {\bf 81}, 5932 (1998).

\bibitem{San11}
Nicolas Sangouard, Christoph Simon, Hugues de Riedmatten, and Nicolas Gisin
Rev. Mod. Phys. {\bf 83}, 33 (2011).

\bibitem{Sre16}
S. Muralidharan, L. Li, J. Kim, N. L\"{u}tkenhaus, M. D. Lukin, and L. Jiang
Scientific Reports {\bf 6}, Article number: 20463 (2016).

\bibitem{Cirac99}
J. I. Cirac, A. K. Ekert, S. F. Huelga, and C. Macchiavello, Phys. Rev. A {\bf 59}, 4249 (1999).

\bibitem{Crep02}
C. Crepeau, D. Gottesman, and A. Smith,
arxiv:quant-ph/0206138 (2002).

\bibitem{Crep06}
M. Ben-Or, C. Crepeau, D. Gottesman,  A. Hassidim, and A. Smith, 
 Proceedings of the 47th Annual IEEE Symposium on Foundations of Computer Science (FOCS'06)
0-7695-2720 (2006).

\bibitem{Komar13}
P. K\'om\'ar, E. M. Kessler, M. Bishof, L. Jiang, A. S. S\o rensen, J. Ye, and M. D. Lukin,
Nature Physics {\bf 10}, 582 (2014).

\bibitem{Gisin}
N. Gisin, G. Ribordy, W. Tittel, and H. Zbinden                                                                    
Rev. Mod. Phys. {\bf 74}, 145 (2002).

\bibitem{And04} 
A. S. S\o{}rensen, C. H. van der Wal, L. I. Childress and M. D. Lukin, 
Phys. Rev. Lett. {\bf 92}, 063601 (2004).

\bibitem{Rab06} 
P. Rabl, D. DeMille, J. M. Doyle, M. D. Lukin, R. J. Schoelkopf and P. Zoller, Phys. Rev. Lett., {\bf 97}, 033003 (2006).

\bibitem{And06}
A. Andre, D. DeMille, J. M. Doyle, M. D. Lukin, S. E. Maxwell, P. Rabl, R. J. Schoelkopf and P. Zoller, 
Nature Physics, {\bf 2}, 636 (2006).

\bibitem{Wall09}
M. Wallquist, K. Hammerer, P. Rabl, M. Lukin, and P. Zoller, Phys. Scr. {\bf T137}, 014001 (2009).

\bibitem{Taylor11}
J. M. Taylor, A. S. S\o rensen, C. M. Marcus, and
E. S. Polzik,  Phys. Rev. Lett. {\bf 107}, 273601 (2011).

\bibitem{vitali12}
S.  Barzanjeh, M. Abdi, G. J. Milburn, P.
Tombesi, and D. Vitali, Phys. Rev. Lett. {\bf 109}, 130503 (2012).

\bibitem{Tian12}
L. Tian, Phys. Rev. Lett. {\bf 108}, 153604 (2012).

\bibitem{clerk12}
Y.-D. Wang and A. A. Clerk, Phys. Rev. Lett. {\bf 108}, 153603 (2012).

\bibitem{Gis13}
Z-L. Xiang, S. Ashhab, J. Q. You, and F. Nori, Rev. Mod. Phys. {\bf 85}, 623 (2013).

\bibitem{Mar10} 
D. Marcos, M. Wubs, J. M. Taylor, R. Aguado, M. D. Lukin and A. S. S\o{}rensen,  
Phys. Rev. Lett., {\bf 105}, 210501 (2010).

\bibitem{Xia14}
K. Xia, M. R. Vanner	and J. Twamley, Sci. Rep. {\bf 4}, 5571 (2014).

\bibitem{Bri14}
C. OÕBrien, N. Lauk, S. Blum, G. Morigi, and M. Fleischhauer
Phys. Rev. Lett. {\bf 113}, 063603 (2014).

\bibitem{Pri14}
J. D. Pritchard, J. A. Isaacs, M. A. Beck, R. McDermott, and M. Saffman
Phys. Rev. A {\bf 89}, 010301 (2014).

\bibitem{Will14}
L. A. Williamson, Y.-H. Chen, and J. J. Longdell,
Phys. Rev. Lett. {\bf 113}, 203601 (2014).

\bibitem{Duan15}
Z. Q. Yin, W. L. Yang, L. Sun, L. M. Duan, Phys. Rev. A {\bf 91}, 012333 (2015).

\bibitem{Ham15}
Ondrej Cernotik and  Kl. Hammerer, arxiv:1512.00768 (2015).

\bibitem{Wall04} 
A. Wallraff, D. I. Schuster, A. Blais, L. Frunzio, R.- S. Huang, J. Majer, S. Kumar, S. M. Girvin and R. J. Schoelkopf, 
Nature {\bf 431},162 (2004).

\bibitem{Gre14}
Y. Kubo, F. R. Ong, P. Bertet, D. Vion, V. Jacques, D. Zheng, A. Dr\'{e}au, J.-F. Roch, A. Auffeves, F. Jelezko, J. Wrachtrup, M. F. Barthe, P. Bergonzo, and D. Esteve,
Phys. Rev. Lett. {\bf 105}, 140502 (2010).

\bibitem{bochmann13}
J. Bochmann, A. Vainsencher, D. D. Awschalom, and A. N. Cleland, Nature Physics {\bf 9}, 712 (2013).

\bibitem{Andrews14}
R. W. Andrews, R. W. Peterson, T. P. Purdy, K. Cicak,
R. W. Simmonds, C. A. Regal, and K. W. Lehnert, Nature Physics {\bf 10}, 321 (2014).

\bibitem{Bagchi14}
T. Bagci, A. Simonsen, S. Schmid, L. G. Villanueva,
E. Zeuthen, J. Appel, J. Taylor, A. S. S\o rensen,
K. Usami, A. Schliesser, and E. S. Polzik, Nature {\bf 507}, 81 (2014).

\bibitem{Nat12}
C. M. Natarajan, M. G. Tanner and R. H. Hadfield, Supercond. Sci. Technol. {\bf 25}, 063001 (2012).

\bibitem{Pro13}
S. Probst, H. Rotzinger, S. W\"{u}nsch, P. Jung, M. Jerger, M. Siegel, A. V. Ustinov, and P. A. Bushev,
Phys. Rev. Lett. {\bf 110}, 157001 (2013).

\bibitem{schu10}
D. I. Schuster, A. P. Sears, E. Ginossar, L. DiCarlo, L. Frunzio, J. J. L. Morton, H. Wu, G. A. D. Briggs, B. B. Buckley, 
D. D. Awschalom, and R. J. Schoelkopf, Phys. Rev. Lett. {\bf 105}, 140501 (2010).

\bibitem{Tor08}
K. Tordrup and K. M\o lmer, Phys. Rev. A {\bf 77}, 020301 (2008).

\bibitem{atac09}
A. Imamo\v{g}lu, Phys. Rev. Lett. {\bf 102}, 083602 (2009).

\bibitem{ver09}
J. Verd\'{u}, H. Zoubi, Ch. Koller, J. Majer, H. Ritsch, and J. Schmiedmayer, Phys. Rev. Lett. {\bf 103}, 043603 (2009).

\bibitem{zhu11}
X. Zhu, S. Saito, A. Kemp,	K. Kakuyanagi,	S. Karimoto, H. Nakano, W. Munro, Y. Tokura, M. S. Everitt, K. Nemoto, M,Kasu, N. Mizuochi and K. Semba,
Nature {\bf 478}, 221Ð224 (2011).

\bibitem{Blum15}
S. Blum, C. O'Brien, N. Lauk, P. Bushev, M. Fleischhauer, and G. Morigi
Phys. Rev. A {\bf 91}, 033834 (2015).

\bibitem{Cat11}
G. Catelani, R. J. Schoelkopf, M. H. Devoret, and L. I. Glazman
Phys. Rev. B {\bf 84}, 064517 (2011).

\bibitem{Orrit92}
M. Orrit, J. Bernard, A. Zumbusch, and R. I. Personov, Chem.
Phys. Lett. {\bf 196}, 595 (1992).

\bibitem{Mich99}
Ch. Brunel, Ph. Tamarat, B. Lounis, J. C. Woehl, and Michel Orrit, J. Phys. Chem. A {\bf 103}, 2429 (1999).

\bibitem{Vahid12}
Y. L. A. Rezus, S. G. Walt, R. Lettow, A. Renn, G. Zumofen, S. G\"{o}tzinger and V. Sandoghdar,  
Phys. Rev. Lett., {\bf 108}, 093601, (2012).

\bibitem{San14}
S. Faez, P. T\"{u}rschmann, H. R. Haakh, S. G\"{o}tzinger and V. Sandoghdar,
Phys. Rev. Lett. {\bf 113}, 213601 (2014).

\bibitem{Gaio16}
M. Gaio, M. Moffa, M. Castro-Lopez, D. Pisignano, A. Camposeo, and R. Sapienza,
ACS Nano {\bf 10}, 6125 (2016).

\bibitem{Skoff16}
S. M. Skoff, D. Papencordt, H. Schauffert, and  Arno Rauschenbeutel, 
arXiv:1604.04259 (2016).

\bibitem{Alm04}
V. R. Almeida, Qianfan Xu, C. A. Barrios and M. Lipson,  
Optics Letters, {\bf 29},1209 (2004).
		
\bibitem{Qua09}
Q. Quan, I. Bulu and M. Lon\'car,  
Phys. Rev. A, {\bf 80}, 011810 (2009).

\bibitem{Mak01} 
Y. Makhlin, G. Schoen and A. Shnirman, Rev. Mod. Phys. {\bf 73}, 357 (2001).

\bibitem{Dev98}
V. Bouchiat, D. Vion, P. Joyez, D. Esteve and M. H. Devoret
Physica Scripta. {\bf T76}, 165, (1998).

\bibitem{Vion02}
D. Vion, A. Aassime, A. Cottet, P. Joyez, H. Pothier, C. Urbina, D. Esteve, M. H. Devoret,
Science  {\bf 296}, 886 (2002).

\bibitem{supp}
See supplementary material [url], which includes Refs. [56 - 58]

\bibitem{Das16}
S. Das, V. Elfving, F. Reiter, and A. S. S\o{}rensen, in preparation

\bibitem{dirk}
D. Witthaut, and A. S. S\o rensen, New Journal of Physics, {\bf12} 043052 (2010).

\bibitem{chang}
Y. Li, L. Aolita, D. E. Chang, and L. C. Kwek, Phys. Rev. Lett. {\bf 109}, 160504 (2012).

\bibitem{book}
M. O. Scully and M. H. Zubairy, \textit{Quantum Optics}, Cambridge University Press (1997).

\bibitem{Blais04}
A. Blais, R.-S. Huang, A. Wallraff, S. M. Girvin, and R. J. Schoelkopf,
Phys. Rev. A {\bf 69}, 062320 (2004).

\bibitem{Wirtz06}
A. C. Wirtz , C. Hofmann , and E. J. J. Groenen, 
J. Phys. Chem. B {\bf 110}, 21623 (2006)

\bibitem{Sanprb14}
S. Faez, S. J. van der Molen, and M. Orrit, Physical Review B, {\bf 90}, 205405 (2014).
				
\bibitem{Ith05}
G. Ithier, E. Collin, P. Joyez, P. J. Meeson, D. Vion, D. Esteve, F. Chiarello, A. Shnirman, Y. Makhlin, J. Schriefl, and G. Sch\"{o}n, 
Phys. Rev. B. {\bf 72}, 134519 (2005).
  
\bibitem{Hou09}
A. A. Houck, J. Koch, M. H. Devoret, S. M. Girvin and R. J. Schoelkopf, Quantum Inf Process {\bf 8}, 105 (2009).

\bibitem{Hett02}
C. Hettich, C. Schmitt, J. Zitzmann, S. KŸhn, I. Gerhardt, V. Sandoghdar,
Science {\bf 298}, 385-389 (2002).

\bibitem{flo12}
F. Reiter and A. S. S\o{}rensen, Phys. Rev. A, {\bf 85}, 032111 (2012).

\bibitem{Das08}
S. Das, G.S. Agarwal, and M. O. Scully, Phys. Rev. Lett {\bf 101}, 153601 (2008).
	
\bibitem{zoll99}
C. Cabrillo, J. I. Cirac, P. Garcia-Fern\'andez and P. Zoller, 
Phys. Rev. A, {\bf 59}, 1025 (1999).

\bibitem{Lillian05}
L. I. Childress, J. M. Taylor, A. S. S\o{}rensen, and M. D. Lukin, Phys. Rev. A {\bf 72}, 052330 (2005).

\bibitem{CHSH1}
J. F. Clauser, M. A. Horne, A. Shimony, and R. A. Holt,
Phys. Rev. Lett. {\bf 23}, 880 (1969).

\bibitem{CHSH2}
A. Aspect, P. Grangier, and G. Roger
Phys. Rev. Lett. {\bf 47}, 460 (1981).

\bibitem{Wall05}
A. Wallraff, D. I. Schuster, A. Blais, L. Frunzio, J. Majer, M. H. Devoret, S. M. Girvin, and R. J. Schoelkopf, Phys. Rev. Lett. {\bf 95}, 060501 (2005).

\end{thebibliography}

\begin{thebibliography}{999}

\bibitem{flo12}
F. Reiter, and A. S. S\o{}rensen, 
Phys. Rev. A, {\bf 85}, 032111 (2012).

\bibitem{Das16}
S. Das, V. Elfving, F. Reiter, and A. S. S\o{}rensen, in preparation

\bibitem{Vion02}
D. Vion, A. Aassime, A. Cottet, P. Joyez, H. Pothier, C. Urbina, D. Esteve, M. H. Devoret,
Science  {\bf 296}, 886 (2002).

\bibitem{dirk}
D. Witthaut, and A. S. S\o rensen, New Journal of Physics, {\bf12} 043052 (2010).

\bibitem{chang}
Y. Li, L. Aolita, D. E. Chang, and L. C. Kwek, Phys. Rev. Lett. {\bf 109}, 160504 (2012).

\bibitem{CHSH1}
J. F. Clauser, M. A. Horne, A. Shimony, and R. A. Holt,
Phys. Rev. Lett. {\bf 23}, 880 (1969).

\bibitem{CHSH2}
A. Aspect, P. Grangier, and G. Roger
Phys. Rev. Lett. {\bf 47}, 460 (1981).

\bibitem{Wall05}
A. Wallraff, D. I. Schuster, A. Blais, L. Frunzio, J. Majer, M. H. Devoret, S. M. Girvin, and R. J. Schoelkopf, 
Phys. Rev. Lett. {\bf 95}, 060501 (2005).

\end{thebibliography}
\end{document}